\documentstyle[12pt]{article}

\newcommand{\inte}{\int_{T_{x_{i}}(C)}d^{2}\xi_{\mu}
\int_{T_{x_{j}}(C)}d^{2}\eta_{\alpha} \partial^{(\xi)}
_{\nu}arg(\vec{\xi}-\vec{x}_{i})
\partial^{(\eta)}_{\beta}arg(\vec{\eta}-\vec{x}_{j})}

\newcommand{\soma}{\sum^2_{i,j=1}\epsilon_i\epsilon_j}

\newcommand{\cors}{<\sigma(x)\sigma^*(y)>}

\newcommand{\corm}{<\mu(x)\mu^*(y)>}

\newcommand{\limrd}{\lim_{\rho ,\delta\rightarrow 0}}

\newcommand{\limrde}{\lim_{\rho ,\delta ,\epsilon\rightarrow 0}}

\newcommand{\tendei}{\stackrel{|x-y|\rightarrow\infty}{\longrightarrow}}

\newcommand{\integ}{\int_{R_{r_i}(C)} d^2\xi \int_{T_{s_j}(C)} d^2\eta
_\alpha [\epsilon^{ik} \partial^{(\xi)}_k arg(\vec \xi-\vec r_i)+
\partial^i_{(\xi)}\ln |\vec \xi -\vec r_i|] }

\newcommand{\limrdxy}{\lim_{\rho_x,\delta_x,\rho_y,\delta_y\rightarrow
0}}

\newcommand{\limrdx}{\lim_{\rho_x,\delta_x\rightarrow 0}}

\newcommand{\integra}{\int_{R_{r_i}(C)} d^2\xi \int_{V(S)} d^3\eta
 [\epsilon^{ik} \partial^{(\xi)}_k arg(\vec \xi-\vec r_i)+
\partial^i_{(\xi)}\ln |\vec \xi -\vec r_i|]}

\begin{document}

\begin{titlepage}

\topmargin -10mm
\rightline{PUPT-1330}
\rightline{August 1992}

\vskip 10mm

\centerline{ \LARGE\bf Duality, Quantum Vortices and Anyons }
\vskip 3pt

\centerline{ \LARGE\bf in Maxwell-Chern-Simons-Higgs Theories }

\vskip 15mm

\centerline{\sc E.C. Marino\footnote{On sabbatical leave from Departamento de
F\'{\i}sica, Pontif\'{\i}cia Universidade Cat\'{o}lica, Rio de Janeiro, Brazil.
E-mail:  marino@puhep1.princeton.edu}}
\vskip 6mm

\centerline{\it Joseph Henry Laboratories}
\centerline{\it Princeton University}
\centerline{\it Princeton, New Jersey 08544}
\vskip 20mm
 \begin{abstract}
The order-disorder duality structure is exploited in order to obtain a
quantum description of anyons and vortices in: a) the Maxwell theory;
b) the Abelian Higgs Model; c) the Maxwell-Chern-Simons theory; d) the
Maxwell-Chern-Simons-Higgs theory.  A careful construction of a charge
bearing order operator($\sigma$) and a magnetic flux bearing
 disorder operator (vortex operator) ($\mu$)
 is performed, paying attention to the necessary requirements
for locality.  An anyon operator is obtained
 as the product $\varphi=\sigma\mu$.  A
detailed and comprehensive study of the euclidean correlation functions
of $\sigma$, $\mu$ and $\varphi$ is carried on in the four theories
above.  The exact correlation functions are obtained in cases
$\underline{a}$ and $\underline{c}$.  The large distance behavior of
them is obtained in cases $\underline{b}$ and $\underline{d}$.  The study
of these correlation functions allows one to draw conclusions about the
condensation of charge and magnetic flux, establishing thereby an analogy
with the Ising model.  The mass of vortex and anyon excitations is
explicitly obtained wherever these excitations are present in the
spectrum.  The independence between the mechanisms of mass generation
for the vortices and for the vector field is clearly exposed.
\end{abstract}

\vfill
\end{titlepage}

\leftmargin 25mm
\topmargin -8mm
\hsize 153mm
\baselineskip 7mm
\setcounter{page}{2}

\newfont{\myfont}{msbm10 scaled\magstep1}

\leftline{\Large\bf 1) Introduction}
\bigskip
The physics of $2+1$ dimensional world has been the object of intense
theoretical investigation in the last few years.  One of the features
which has aroused a lot of interest is the possibility of existence of
states with generalized, continuous statistics.  Chern-Simons field
theories [1] have been playing a central role in this framework.  An
important reason for this is that the presence of a Chern-Simons term in
the lagrangian of a vector field induces a change in the statistics of
the charged particles eventually coupled to it [2].  This statistical
transmutation effectively occurs because the coupling of a charged
particle to a Chern-Simons vector field imparts to this particle a point
magnetic flux [2] and, as was demonstrated in full generality [3], the
bound state of a point charge and a point magnetic flux carries
arbitrary statistics proportional to the product: charge $\times$  magnetic
flux.
A particular case of this fact was first observed in a specific model in
[4].

The mechanism of statistical transmutation via the Chern-Simons
lagrangian is a key ingredient of field theories proposed to describe
very interesting condensed matter systems such as the two dimensional
electron gas undergoing the Quantum Hall effect [5].  It was also
invoked in the theory of superconductivity [6] involving particles with
generalized statistics or anyons as they become known.

In view of their vast potential physical interest, a large amount of
study of anyon properties was undergone recently [7].  In order to
obtain a full quantum field theoretic description of anyons one should
be able to evaluate their correlation functions, mass spectrum and
scattering amplitudes.  The purpose of this work is to accomplish some
of these goals in the framework of four different quantum field
theories: a) the Maxwell theory; b) the Abelian Higgs Model
(Maxwell-Higgs theory); c) the Maxwell-Chern-Simons theory; d) the
Maxwell-Chern-Simons-Higgs theory.

A lot of work has been devoted to the study of generalized statistics in
$1+1$ D [8,4,9].  Actually the investigation performed in $1+1$D
provided a fundamental clue for understanding the quantum description of
anyons, either in $1+1$D and $2+1$D.  This consists in
 the realization that there
is a basic algebraic structure underlying the existence of generalized
statistics excitations, namely, the order-disorder duality [10].  The
idea is to introduce order ($\sigma$) and disorder ($\mu$) operators
satisfying a certain algebra (dual algebra) in such a way that the
composite operator $\varphi=\sigma\mu$ made out of them is shown to be
in general an anyon operator [4].  One then generalizes the method of
Kadanoff and Ceva [10] for continuum field theory [4,11] in order to
obtain the euclidean correlation functions of $\sigma$, $\mu$ and
$\varphi$.  A review of this method is in ref. [12].   Very interesting
and important related works are found in [13].

A few years ago, the order-disorder duality structure was generalized
for $2+1$-dimensional continuum field theory [14].  A magnetic flux
bearing disorder operator $\mu$ was introduced and its correlation
functions obtained as euclidean functional integrals.  It was also shown
that the order operator $\sigma$ dual to $\mu$ should now be a charge
bearing operator.  As a consequence, the product $\sigma\mu$ is
naturally seen to be an anyon operator.  This description of anyons
goes beyond the Chern-Simons mechanism because one can study generalized
statistics excitations even in theories without a Chern-Simons term.
Very interesting related works were done previously on the the lattice
[15] and subsequently in the continuum [16].

An extremely interesting and appealing feature of this approach emerges
when one realizes that the magnetic flux is the topological charge in
$2+1$D.  Hence the disorder operator $\mu$ or the anyon itself, must be
closely related to topological charge bearing excitations in $2+1$D,
namely, topological solitons or vortices.  This relationship was
investigated and demonstrated in [14,16] where the disorder variable
$\mu$ was shown to be a quantum vortex creation operator.

The above ideas were recently applied in the complete bosonization of
the Dirac fermion field in $2+1$D [18]. In this case the $\sigma$ and
$\mu$ dual operators were expressed in terms of a bosonic vector field
$W_\mu$ with dynamics governed by a nonlocal version of the Maxwell
action plus a Chern-Simons term.

It follows from the work done in [14] that some stringent requirements
must be met in order for the vortex operator to be local.  Also, a lot
of care must be exercised in order to obtain the charge bearing local
order operator $\sigma$ dual to it.  We will carefully analyze the
construction of local order, disorder and anyon operators in the various
theories under consideration, according to the requirements set forth in
[14].

After expressing the fields $\sigma$, $\mu$ and $\varphi$ in terms of a
vector field $W_\mu$, we make a detailed study of the euclidean
correlation functions of these operators and evaluate the commutation
rules among themselves as well as between them and the charge and
magnetic flux operators in the four theories mentioned above.  Exact
results for all correlation functions are obtained in the case of the
Maxwell and Maxwell-Chern-Simons theories.  The long distance behavior
of the correlation functions is obtained in the case of the Abelian
Higgs Model and Maxwell-Chern-Simons-Higgs theories, both in the
symmetric and broken phases.

We investigate the conditions which allow the presence of genuine vortex
on anyon excitations in the spectrum and in these cases, evaluate their
masses, for all of the above mentioned theories.

As in $1+1$D [4], we observe that a nontrivial commutation relation
between operators, manifests itself in the multivaluedness of the
corresponding euclidean correlation functions.  Analytic continuation
back to Minkowski space, from each sheet, will lead to the various
orderings of operators.  We are thereby able to retrieve the commutation
rules directly from the study of euclidean correlation functions. We
show that the general expression for the statistics of a field
bearing a charge Q and a magnetic flux $\Phi$ is S=$\frac{\tilde Q\Phi}{
4\pi}$ , where $\tilde{Q}=Q-\theta\Phi$ with $\theta$ being the
coefficient of the Chern-Simons term eventually present in the theory.

We can envisage some interesting applications of our formalism in field
theories related to condensed matter systems.  Among them we could
mention the Chern-Simons-Landau-Ginsburg theory of the Quantum Hall
Effect [19] and the theories describing the Hall-Superconducting phase
transition in anyon systems [20].  We are presently investigating such
applications.

The material is organized in the next four chapters in such a way that
each chapter corresponds to one of the theories mentioned above.
Concluding remarks and the main conclusions are presented in chapter 6.
Six Appendixes are included to demonstrate useful results.
\vfill
\newpage
\leftline{ \Large\bf 2) The Maxwell Theory}
\bigskip
\leftline{\large\bf 2.1) Order and Disorder(Vortex) Operators}
\bigskip
\leftline{\bf 2.1.1) The Disorder (Vortex) Operator}
Let us start by considering the Maxwell theory, described by
$${\cal L} = - {1\over 4} W_{\mu\nu} W^{\mu\nu}\eqno(2.1)$$
which leads to Maxwell equation
$$\partial_\nu W^{\nu\mu}=0\eqno(2.2)$$
The identically conserved topological current is given by
$$J^\mu =\epsilon^{\mu\alpha\beta}\partial_\alpha W_\beta\eqno(2.3)$$
We can also introduce the two indexes dual topological current
$$\tilde J^{\mu\nu} \equiv \epsilon^{\mu\nu\alpha} J_\alpha
\,\,\,\,\,\,\,\,\,\,\, or\,\,\,\,\,\,\,\,\,\,\,
\tilde J^{\mu\nu} = W^{\mu\nu}\eqno(2.4)$$
$\tilde J^{\mu\nu}$ is not identically conserved; rather, its
conservation is implied by the field equation (2.2).

A topological charge (magnetic flux) bearing operator (vortex operator)
was first introduced as a disorder variable in continuum $2+1$ D field
theory in [14].  A fundamental piece in the construction of the vortex
operator in [14] and which was also used in [18] is the singular
external field
$$A_\mu (z;x;C;T_x) = \int_{ T_x(C)} d^2\xi_\mu arg(\vec \xi -\vec x)
\delta^3 (z-\xi)\eqno(2.5)$$
In this expression, $arg(\vec z)$ is the angle the vector $\vec z$ makes
with the $z^1$ axis $(z=(z^0,\vec z))$. $T_x(C)$ is the portion of the
 ${\myfont R}^2$  plane at $\xi^0=x^0$, external to the curve C.
C is the curve
depicted in Fig. 1 which contains the arc of a circumference of radius
$\rho$, centered on $(x^0,\vec x)$ and two straight lines along the cut of
$arg(\vec z -\vec x)$ separated an angle $2\delta$ apart.

The vortex operator (disorder operator) $\mu$ is one of the two basic
dual operators whose properties we are going to investigate in this
work.  It is obtained by coupling the $A_\mu$ external field to the dual
current $\tilde J^{\mu\nu}= W^{\mu\nu}$ in the following way
[14]
$$\mu(x;C)=exp\lbrace -{{ib}\over 2}\int d^3z W^{\mu\nu}
A_{\mu\nu}\rbrace\eqno(2.6)$$
In this expression, b is a free real parameter with dimension of
$(mass)^{- {1\over 2}}$ and $A_{\mu\nu} =\partial_\mu A_\nu
-\partial _\nu A_\mu$. In [14] it is shown that the correlation
functions of $\mu$ are completely independent of the surface $T_x$
appearing in the definition of $A_\mu$, provided we introduce a surface
renormalization counterterm depending on $A_\mu$ and having the same
form as the kinetic term of the $W_\mu$ field, namely
$${\cal L}_R = -{b^2\over 4} A_{\mu\nu} A^{\mu\nu}\eqno(2.7)$$

In the presence of this counterterm, it is clear that the integrand in
the functional integral defining $\mu$ correlation functions depends on
the external field $A_\mu$ through the conbination $W_\mu +A_\mu$.( See
Eq.(2.26)) This is a fundamental requirement for surface invariance
because making a change of variable of the type $W_\mu\rightarrow W'_\mu
= W_\mu +\partial_\mu \omega$, which of course leaves the measure invariant
we have $W_\mu + A_\mu \rightarrow W'_\mu +A_\mu -\partial_\mu \omega$.  One
can demonstrate [14] that for an appropriate choice of $\omega$, we get
$A_\mu(z;x;C;T_x)-\partial_\mu \omega =A_\mu(z;x;C;\tilde T_x)$, where
$$A_\mu(z;x;C;\tilde T_x)=\int_{\tilde T_x} d^2\xi_\mu arg(\vec \xi -\vec
x)\delta^3(z-\xi) -\Theta (V(\tilde T_x))\partial_\mu arg(\vec z-\vec
x)\eqno(2.8)$$
Here $\tilde T_x$ is an arbitrary surface bounded by C and $V(\tilde
T_x)$ is the volume enclosed by $\tilde T_x(C)$ and $T_x(C)$.
$\Theta(V(\tilde T_x))$ is the three-dimensional heaviside function with
support on $V(\tilde T_x)$. (See Fig. 2 for $\tilde T_x(C)$ and
$V(\tilde T_x)$).  It is not difficult to realize [14] that the $\omega$ which
leads from (2.5) to (2.8) is
$$\omega =\Theta (V(\tilde T_x)) arg (\vec z-\vec x)\eqno(2.9)$$
$A_\mu (z;x;C; \tilde T_x)$, eq. (2.8), is the most general form of the
external field [14].  Observe that $A_\mu (z;x;C;T_x)$, eq. (2.5), is a
special case of $A_\mu(z;x;C;\tilde T_x)$, eq. (2.8); when $\tilde T_x$
reduces to $T_x$, since, in this case $V(\tilde T_x)$ goes to zero along
with the second term in (2.8).  As one would expect, the external field
intensity tensor is not changed by the transformation which takes (2.5)
into (2.8), namely $A_{\mu\nu} [A_\mu(T_x)] =A_{\mu\nu}[A_\mu(\tilde
T_x)]$.  We emphasize that in the presence of the counterterm (2.7), as
we will explicitly see, the correlation functions of $\mu$ are
completely independent of $T_x$ or $\tilde T_x$, only depending on x and
C.  A local limit for $\mu$ can then be obtained by using $\rho$, the
radius of the circumference part of C and $\delta$, the angular width
of the region along the cut (See Fig. 1), as regulators and at the end
of all calculations, taking the limit in which $\rho$ and $\delta$
vanish.  As we will see, the $\mu$ correlation functions will only depend
on x in this limit.  This is the procedure we are going to adopt
in order to obtain local vortex (disorder) correlation functions and
operators.

\bigskip
\leftline{\bf 2.1.2) The Order Operator}

Let us introduce here the order operator which is going to be dual to
$\mu$.  The natural way of defining the order operator, which is
suggested by the study of a wide class of models in $1+1D$ [12] and also
in $2+1D$ [18,24], is to couple the external field $A_\mu$ to the
topological current $J^\mu$ in the following way
 $$\Sigma(x;C) =\exp
\lbrace ia\int d^3z J^\mu A_\mu\rbrace = \exp\lbrace ia\int d^3z
\epsilon ^{\mu\alpha\beta} A_\mu \partial _\alpha W_\beta \rbrace\eqno
(2.10a)$$
or
$$\Sigma'(x;C) = \exp \lbrace ia \int d^3z W^\mu J_\mu[A_\mu]\rbrace =
\exp\lbrace ia \int d^3z \epsilon^{\mu\alpha\beta} W_\mu
\partial_\alpha A_\beta \rbrace\eqno(2.10b)$$
which differs from $\Sigma$ by a boundary term.  In (2.10), \underline{a}
is a free real parameter with dimension of $(mass)^{1\over 2}$.

It turns out that we cannot obtain a local operator out of $\Sigma$ or
$\Sigma'$ even in the limit where $\rho$,$\delta\rightarrow 0$.  No
surface renormalization counterterm will render the $\Sigma$
correlation functions surface invariant in theories containing a Maxwell
term in the action.  This is so, because in contraposition to the case
of $\mu$, no renormalization counterterm will ever be able to make the
integrand in the functional integral describing $\Sigma$ or $\Sigma'$
correlation functions to depend on the external field $A_\mu$ in the
combination $W_\mu +A_\mu$, such that the procedure employed in [14] to
show the surface invariance of $\mu$ could be applied.  The only case in
which the $\Sigma$ or $\Sigma '$ operators could be made local would be
a theory containing just the Chern-Simons term in its kinetic
lagrangian.  In this case the renormalization counterterm would be a
Chern-Simons term involving the external field.  In this case however,
the $\mu$ operator would no longer be local!  Actually, as we will see,
in theories containing both the Maxwell and Chern-Simons terms, we will
have to redefine $\mu$ by adding a $\Sigma'$ piece, in order to obtain a
local disorder (vortex) operator.

In order to construct a local operator $\sigma(x)$ dual to $\mu(x;C)$,
we are going to take advantage of the Cauchy-Riemann equations relating
the real and imaginary parts of the analytic function  ln z(see Appendix
A).  Let us define
$$\sigma(x^0,\vec x) = \lim_{\rho,\delta\rightarrow 0} \exp \lbrace ia
\int _{R_x(C)} d^2 \xi [\epsilon^{ij}\partial ^{(\xi)}_i arg (\vec \xi
-\vec x) + \partial ^{(\xi)}_j \ln \vert \vec \xi - \vec x\vert] W_j
(x^0, \vec \xi)]\rbrace\eqno(2.11)$$

In the above expression \underline{a} is a free real parameter with dimension
of
$(mass)^{1\over 2}$ and the integral is performed over the region
$R_x(C)$, depicted in Fig. 3.  $R_x(C)$ is the part of the ${\myfont R}^2$
 plane
external to the curve $C$.  This consists of the arc of a circumference
of radius $\rho$ centered on $\vec x$ and crossing the cut of the
function arg $(\vec \xi -\vec x)$ and two straight lines, along this cut,
separated an angle $2\delta$ apart.

Using the Cauchy-Riemann equation, eq.(A.2) one immediately realizes
that the exponent in (2.11) is different from zero only on the
singularities of the arg$(\vec \xi -\vec x)$ and ln $\vert \vec \xi
-\vec x\vert$ functions, that is, the point $x =(x^0, \vec x)$ and the
cut of arg$(\vec \xi -\vec x)$.  It is therefore clear that when we take
out the regulators: $\rho, \delta \rightarrow 0$, $\sigma$ will only
depend on $x$.  This will be made at the end of all calculations, as in
the case of $\mu$.

As we shall see, $\sigma$ as given by (2.11) is the correct local
operator dual to $\mu$.

\bigskip
\leftline{ \large\bf 2.2)Commutation Rules. Anyon Operators}

Let us study here the various commutation rules involving the $\sigma$
and $\mu$ operators.  These will enable us to construct an anyon
operator out of $\sigma$ and $\mu$.

The basic equal-time commutators of the Maxwell theory are
$$[W^i, W^j]=[E^i,E^j]=[\Pi^i,\Pi^j]=0\qquad\qquad
i,j=1,2$$
$$
[W^i(x),E^j(y)]= -[W^i(x),\Pi^j(y)]= -i \delta^{ij}\delta^2 (\vec
x-\vec y)\}\eqno(2.12)$$

Here $E^i=W^{io}$ is the electric field and $\Pi^i=-E^i$ is the momentum
canonically conjugate to $W^i$.

We show in Appendix B that the $\mu$ operator, eq. (2.6) can be cast in
the form
$$\mu(x;C)= \exp \lbrace -ib \int_{T_x (c)} d^2\xi E^i
(x^0,\vec\xi)\partial_i arg(\vec \xi -\vec x)\rbrace\eqno(2.13)$$

Using (2.12) we immediately see that $[\mu,\mu] = [\sigma\sigma] =0$.
The $\mu -\sigma$ commutation rule can be obtained by writing
$\mu_{\rho\delta} (x;C) \equiv e^{A(x,C)},
\sigma_{\rho\delta}(y;C)\equiv e^{B(y,C)}$
and using the formula $e^Ae^B=e^Be^Ae^{[A,B]}$, which is valid when
$[A,B]$ is a c-number.  Making use of (2.13), (2.11) and (2.12), we
find
$$[A(x;C), B(y;C)]=ab\int_{T_x(C)} d^2\xi \int_{R_y(C)}d^2\eta
\partial_i  arg(\vec \xi-\vec x)[\epsilon^{kl}\partial_k arg (\vec \eta
-\vec y)$$
$$
 +\partial_l \ln\vert \vec\eta -\vec y\vert]
(-i)\delta^{il}\delta^2 (\vec \xi -\vec \eta)\eqno(2.14a)
$$
or
$$
[A(x;C), B(y;C)] = -iab\int_{R_y (C)\cap T_x(C)} d^2\eta [\epsilon ^{kl}
\partial_k arg (\vec \eta -\vec y) +\partial_l \ln \vert\vec \eta -\vec
y\vert] \partial_l arg (\vec \eta -\vec x)\eqno(2.14b)$$

In Appendix C it is shown that after taking the limit $\rho ,\delta
\rightarrow 0$ in $R_y(C)$, we obtain the following result for (2.14):
$(B(y)\equiv \lim_{\rho,\delta\rightarrow 0} B(y;C))$
$$[A(x;C), B(y)] =\cases{-i 2\pi ab\ \ arg(\vec y -\vec x)&$ \vec y \in
T_x(C)$\cr 0&$ \vec y \not\in T_x(C)$\cr}\eqno(2.15)$$

We therefore conclude that the equal-time commutation rule between
$\sigma$ and $\mu$ is $(\sigma(y) \equiv \lim_{\rho,\delta \rightarrow 0}
\sigma _{\rho\delta} (y;C))$

$$\mu_{\rho\delta}(x;C)\sigma(y)=\cases {\sigma(y) \mu_{\rho\delta}(x;C)
\exp \lbrace -i 2 \pi ab\ \ arg(\vec y -\vec x)\rbrace &$\vec y \in
T_x(C)$\cr
\sigma (y)\mu_{\rho\delta} (x;C) &$ \vec y \not\in T_x(C)$\cr}\eqno(2.16)$$

This is precisely the U(1) dual algebra appropriate for a vortex
creation operator introduced in [14].  Taking the local limit for
$\mu_{\rho\delta}$, we obtain the local dual algebra (equal times)
$(\mu (x) \equiv \lim _{\rho,\delta \rightarrow 0} \mu_{\rho\delta} (x;C))$

$$\mu(x)\sigma(y) = \sigma(y) \mu(x) exp\lbrace -i 2\pi ab\ \ arg (\vec
y-\vec x)\rbrace\eqno(2.17)$$

Let us investigate now the commutation rules of $\sigma$ and $\mu$ with
the charge and magnetic flux operators, given respectively by

$$Q = \int d^2x\rho(x)\ \ {\rm and}\ \ \Phi = \int d^2x B(x)$$
where
$$
\rho(x)= \partial_i E^i(x)\ \ {\rm and}\ \  B(x) = -\epsilon^{ij}
\partial_i W_j(x)\eqno(2.18)$$

Using (2.12) we readily conclude that $[\sigma, B] =[\mu,\rho] = 0$,
indicating that $\sigma$ does not bear magnetic flux and $\mu$ does not
bear charge.  The commutation rules $[\sigma,\rho]$ and $[\mu,B]$ may be
obtained by the use of the formula $[e^\alpha,\beta] =
e^\alpha[\alpha,\beta]$, valid when $[\alpha,\beta]$ is a c-number.

For $e^{\alpha(x)}\equiv\sigma(x)$ and $\beta(y)\equiv\rho(y)$, using
(2.11), (2.18) and (2.12), we immediately get
$$[\alpha(x),\beta(y)] = \lim_{\rho,\delta\rightarrow 0} (-a) \int
_{R_x(C)} d^2 \xi\, [\epsilon^{ij}\partial_i  arg(\vec \xi -\vec x)
+\partial_j \lim \vert \vec \xi -\vec x \vert] \partial^{(y)}_j \delta^2
(\vec \xi -\vec y)\eqno(2.19a)$$
Using the result (C.1) of Appendix C, and the fact that $\partial^{(y)}
\delta (\xi -y) = -\partial^{(\xi)}\delta(\xi - y)$, we obtain
$$[\alpha(x),\beta (y)] =  2\pi a\, \delta^2 (\vec x -\vec
y)\eqno(2.19b)$$
We therefore have $$[\sigma(x),\rho(y)] =  2 \pi
a\sigma(x)\delta^2(\vec x- \vec y)$$
or
$$[[\sigma(x),Q]=  2 \pi a \sigma (x) Q$$
and
$$[\sigma(x),\Phi] = 0\eqno(2.20)$$
This result indicates that $\sigma(x)$ creates states bearing $
2\pi a$ units of electric charge.

Let us choose now $\epsilon^{\alpha(x)} \equiv \mu (x)$ and
$\beta(y)\equiv B(y)$.  Using (2.13), (2.18) and (2.12), we get
$$[\alpha(x),\beta(y)] = \lim _{\rho,\delta \rightarrow 0} (-b) \int
_{T_x(C)} d^2 \xi\,\, \epsilon^{ij} \partial _j arg(\vec \xi -\vec x)
\partial_i \delta^2 (\vec \xi -\vec y)\eqno(2.21a)$$
Using the result (C.2) of Appendix C, we obtain
$$[\alpha(x),\beta (y)] = 2 \pi b\,\, \delta^2 (\vec x -\vec
y)\eqno(2.21b)$$
We therefore have, in the case of $\mu$ :
$$[\mu(x), B(y)] =2 \pi b\, \mu (x)\,\, \delta^2(\vec x -\vec y)$$
or
$$[\mu(x),\Phi ] = 2\pi b \mu(x)$$
and
$$[\mu(x),Q] = 0\eqno(2.22)$$

This result indicates that $\mu(x)$ creates, states bearing $2\pi b$
units of magnetic flux (or (-) topological charge).

Since $\sigma$ bears charge and $\mu$ bears magnetic flux one should
expect that the composite operator
$$\varphi (x) = \lim_{x\rightarrow y} \sigma(x) \mu (y)
Z(x-y)\eqno(2.23)$$
$-$where $Z(x-y)$ is some renormalization factor used to absorb the short
distance singularities in the above operator product$-$ is an operator
with generalized statistics or an anyon operator.  Indeed, using (2.23)
and (2.17) one immediately finds the equal-times commutation rule for
$\varphi$:

$$\varphi (x)\varphi(y) =\varphi(y)\varphi(x) \exp \lbrace i 2\pi
ab[arg(\vec x -\vec y) -arg (\vec y -\vec x)]\rbrace$$
or
$$\varphi(x) \varphi(y) =\varphi(y)\varphi(x) \exp\lbrace i 2\pi ab
\pi\,\,\,\epsilon[arg(\vec x -\vec y) -\pi]\rbrace\eqno(2.24)$$
where $\epsilon(x)\equiv sign(x)$.  Eq. (2.24) is the appropriate equal
time commutation rule for anyon fields which generalizes the
one-dimensional one [8,4]. Eq. (2.24) indicates that the spin of the
anyon field $\varphi$ is $S_\varphi =\pi ab$, that is, $S_\varphi
={Q\Phi\over 4\pi}$, in agreement with the observations made in [3,4].

Let us remark at this point that the charge bearing operator $\sigma
(x)$, dual to $\mu(x)$ is, of course, gauge noninvariant.  $\mu(x)$ is
essentially the generator of a gauge transformation with parameter
$arg(\vec y -\vec x)$, as can be clearly seen from (2.17) and (2.16).  A
gauge invariant operator would commute with $\mu(x)$ and therefore could
never possibly be its dual!  Indeed, starting from (2.11) and using
(C.1) we see that under the gauge transformation $W_\mu\rightarrow W_\mu
+\Lambda$, we have $\sigma(x)\rightarrow \exp\lbrace i 2\pi
a\Lambda(x)\rbrace\sigma(x)$.  With respect to gauge transformations and
charge bearing, the order operator $\sigma(x)$ is very much like the
scalar field $\phi$ of the abelian Higgs model or the electron field of
Quantum Electrodynamics.

\bigskip
\leftline{\large \bf 2.3) Disorder (Vortex) Correlation Function}

Let us evaluate here the two-point correlation function of $\mu$ within
the euclidean functional integral framework.  Going to the euclidean
space $(ix^0\rightarrow x^3, iW^0\rightarrow W^3, iA^0\rightarrow A^3)$,
we will have
$$iS=i\int d^3z\lbrace -{1\over 4} W_{\mu\nu} W^{\mu\nu}\rbrace
\rightarrow - \int d^3 z_E {1\over 4} W_{\mu\nu} W^{\mu\nu} =$$
$$
-\int d^3 z_E {1\over 2} W_\mu P^{\mu\nu} W_\nu = -S_E\eqno(2.25a)$$
where $P^{\mu\nu} = -\Box_E\delta^{\mu\nu} +
\partial^\mu\partial^\nu$.  The exponent of $\mu$ in(2.6) will
transform as
$$-{ib\over 2} \int d^3z W^{\mu\nu} A_{\mu\nu}\rightarrow
-{b\over 2}\int d^3z_E W^{\mu\nu} A_{\mu\nu} = -b\int d^3 z_E
W_\mu P^{\mu\nu} A_\nu\eqno(2.25b)$$
We will also use a gauge fixing term of the form
$$iS_{GF} = i \int d^3 z[-{\xi\over 2}(\partial_\mu W^\mu)^2]\rightarrow
-{\xi\over 2}\int d^3 z_E(\partial_\mu W^\mu)^2 =$$  $$ -{1\over 2}\int d^3
z_E W_\mu G^{\mu\nu}W_\nu = -S_{E,GF}\eqno(2.25c)$$ where
$G^{\mu\nu} = - \xi \partial^\mu \partial^\nu$ and $\xi$ is an
arbitrary real parameter.  The renormalization counterterm (2.7) transforms
as $$iS_R=i\int d^3 z [-{b^2\over 4}
A_{\mu\nu}A^{\mu\nu}]\rightarrow -\int d^3 z_E{b^2\over
4}A_{\mu\nu} A^{\mu\nu} = $$ $$-\int d^3 z_E {b^2\over 2} A_\mu
P^{\mu\nu} A_\nu =- S_{E,R}\eqno(2.25d)$$

Using (2.25) we can write the following expression for the euclidean
two-point correlation function $<\mu (x) \mu^* (y)>$ (we henceforth will
drop the subscript E):
$$<\mu(x)\mu^*(y)>=\lim_{\rho,\delta\rightarrow 0} Z^{-1}\int D W_\mu
\exp\lbrace -\int d^3 z[{1\over 4} W^2_{\mu\nu} + $$ $${1\over 2} W_\mu
G^{\mu\nu}W_\nu + W_{\mu\nu} A^{\mu\nu} (z;x,y)
+{1\over 4} A^2 _{\mu\nu}(z;x,y)]\rbrace\eqno(2.26a)$$
$$<\mu(x)\mu^*(y)>=\lim_{\rho,\delta \rightarrow 0} Z^{-1}\int D W_\mu
\exp\lbrace -\int d^3 z[{1\over 2} W_\mu [P^{\mu\nu} +
G^{\mu\nu}] W_\nu $$ $$+ W_\mu P^{\mu\nu} A_\nu(z;x,y) +
{1\over 4} A^2_{\mu\nu} (z;x,y)]\rbrace\eqno(2.26b)$$
In these expressions, Z is the vacuum functional and $A_\mu(z;x,y)\equiv
b[A_\mu(z;x;C;T_x)-A_\mu(z;y;C;T_y)]$ (the minus sign corresponds to the
fact that we have the adjoint operator $\mu^*(y)$).  The mixed
$W_\mu-A_\mu$ terms in the exponent in (2.26) come from the
$\mu(x)\mu^*(y)$ operators and the last term is the surface
renormalization counterterm.

Observe that, as we put forward above, the integrand in (2.26) depends
on the external field in the combination $W_\mu + A_\mu$.  Under the
change of variable $W_\mu\rightarrow W_\mu + \partial_\mu \omega$, with
$\omega$ given by (2.9), we can arbitrarily change the surfaces in $A_\mu$
[14].  The gauge fixing term would be invariant under this transformation,
provided we add the ghost term to (2.26) [21].  For simplicity reasons,
however, we will neglect the ghost term in the abelian theories considered
here.  The functional integral in (2.26) can be evaluated with the help of the
euclidean correlation function $<W_\mu W_\nu> =[P^{\mu\nu} + G^{\mu\nu}]^{-1}$.
 This is given, in momentum space, by $$<W^\mu(k)W^\nu
(-k)>={P^{\mu\nu}(k)\over k^4} + {1\over\xi} {k^\mu k^\nu\over k^4}\qquad ;
\qquad P^{\mu\nu}(k) = \delta^{\mu\nu} k^2-k^\mu k^\nu\eqno(2.27a)$$ and in
coordinate space by $$<W^\mu(x)W^\nu(y)> =\lim_{m\rightarrow
0} [P^{\mu\nu} -{1\over \xi} \partial^\mu\partial^\nu]
[{1\over m} -{\vert x- y\vert\over 8\pi}]\ ; \ P^{\mu\nu}
= - \delta^{\mu\nu} +\partial^\mu
\partial^\nu\eqno(2.27b)$$

Here m is an infrared regulator used to control the singularities of
the inverse Fourier transform of ${1\over k^4}$ (see(4.15)).

Using the results of Appendix B, we conclude that each piece of the
second term in the exponent in (2.26b) can be put in the form
$$-b \int d^3z W_\mu P^{\mu\nu} A_\nu = -{b\over 2} \int
d^3z W^{\mu\nu} \partial_\mu A_\nu = \int d^3z B_\gamma
W^\gamma,\eqno(2.28a)$$
where
$$B_\gamma (z;x)= b \int_{T_x(C)} d^2\xi_\alpha\,\, \partial^{(\xi)}_\beta\,
arg(\vec \xi -\vec x) F^{\alpha\beta}\ _\gamma\,\,
\delta^3(z-\xi),\eqno(2.28b)$$ in which $$F^{\alpha\beta}\ _\gamma =
\partial^\alpha \delta^\beta\ _\gamma -\partial^\beta \delta^\alpha\
_\gamma$$

Performing the functional integral in (2.26), we get
  $$<\mu(x)\mu^*(y)>= \lim_{\rho,\delta\rightarrow 0} \exp\lbrace {1\over
2}\int d^3 z d^3 z' B_\mu (z;x,y) B_\nu (z';x,y) $$ $$\times [P^{\mu\nu}
+ G^{\mu\nu}]^{-1}(z-z') -S_R[A_{\mu\nu}]\rbrace\eqno(2.29)$$
where $B_\mu (z;x,y) = B_\mu(z;x)-B_\mu (z;y)$ and
$S_R[A_\mu]$ is the last term in the exponent in (2.26).  Using (2.28b),
(2.27b) and integrating over $z$ and $z'$, we obtain
$$
<\mu(x)\mu^*(y)>= \lim_{\rho,\delta,m\rightarrow 0} \exp
\lbrace {b^2\over 2} \sum^2_{i,j =1} \epsilon_i\epsilon_j \int
_{T_{x_i(C)}} d^2 \xi _\mu \int _{T_{x_j (C)}} d^2 \eta_\alpha
\partial^{(\xi)} _\nu arg(\vec \xi -\vec x_i)\partial^{(\eta)}_\beta arg
(\vec \eta - \vec x _j)$$ $$\times
 F^{\mu\nu} _{(\xi)}\  _\sigma F^{\alpha\beta}_{(\eta)}\ _
\lambda[P^{\sigma\lambda}_{(\xi)} - {1\over \xi} \partial^\sigma
_{(\xi)} \partial^\lambda _{(\xi)}] [{1\over m}
-{\vert\xi-\eta\vert\over 8\pi}] -S_R[A_\mu]\rbrace\eqno(2.30)$$

In the above expression, $x_1\equiv x, \epsilon_1\equiv 1$ and
$x_2\equiv y, \epsilon_2\equiv -1$.

We immediately see that, since
$F^{\alpha\beta}\ _\lambda\partial^\lambda\equiv 0$, the gauge dependent
part of $\ \ \ <W_\mu W_\nu>$ gives no contribution to $<\mu\mu^*>$, as one
should expect for a gauge invariant operator as $\mu$.  The only
contribution comes from the gauge independent part of $<W_\mu W_\nu>$,
which is proportional to $-\Box\delta^{\mu\nu}$.

Let us now make use of the very useful identity
$$
-\Box F^{\mu\nu}_{(\xi)}\ _{\sigma} F^{\alpha\beta}_{(\eta)}\ _{\lambda}
\delta^{\sigma\lambda}=-\Box\epsilon^{\mu\nu\lambda}\epsilon^{\alpha
\beta\gamma}(\partial^{\rho}_{(\xi)}\partial^{\rho}_{(\eta)}\delta
^{\lambda\gamma}-\partial^{\lambda}_{(\xi)}\partial^{\gamma}_{(\eta)})
\eqno(2.31)
$$

which can be readily demonstrated by expanding the $\epsilon $' s in
terms of Kronecker $\delta$'s.  Inserting (2.31) in (2.30) and
observing that
$$
-\Box [\frac{1}{m} - \frac{|x|}{8\pi}]\equiv -\Box {\cal F}^{-1}[\frac{1}
{k^{4}}]={\cal F}^{-1}[\frac{1}{k^{2}}]=\lim_{\epsilon \rightarrow
0} \frac{1}{4\pi [|x|^{2}+|\epsilon |^{2}]^{1/2}}
\eqno(2.32)
$$
(here $\epsilon$ is an ultraviolet regulator used to control the short
distance singularities of the inverse Fourier transform of ${1\over
k^2}$) we get
$$
<\mu (x)\mu^{*}(y)>=\lim_{\rho ,\delta , \epsilon \rightarrow 0}
\exp \{-\frac{b^{2}}{2} \sum_{i,j=1}^{2} \epsilon_{i}\epsilon_{j}
$$
$$\times
\int _{T_{x_{i}(C)}} d^{2}\xi_{\mu} \int_{T_{x_{j}(C)}} d^{2}\eta
_{\alpha} \partial^{(\xi)}_{\nu}arg(\vec{\xi}-\vec{x}_{i})
\partial^{(\eta)}_{\beta}arg(\vec{\eta}-\vec{x}_{j})$$
$$\times\epsilon^{\mu \nu \lambda}\epsilon^{\alpha \beta\gamma}
(\Box_{(\xi)}\delta^{\lambda\gamma}+\partial^{\lambda}_{(\xi)}
\partial^{\gamma}_{(\eta)})[\frac{1}{4\pi [|\xi -\eta |^{2}+
|\epsilon |^{2}]^{1/2}}]-S_{R}[A_\mu]\}
\eqno(2.33)
$$
Here we used the fact that
$\partial_{(\eta)}F(\xi-\eta)=-\partial_{(\xi)} F(\xi-\eta)$.

Observing that $\lim_{\epsilon \rightarrow 0} -\Box [4\pi[\vert x\vert^2
+\vert \epsilon \vert^2]]^{-{1\over 2}} =\delta^3(x)$ and using the
results of Appendix B, we see that the first term in the exponent in
(2.33) can be written as
$$
\frac{b^{2}}{2}\sum_{i,j=1}^{2}\inte  \epsilon^{\mu \nu\lambda}
\epsilon^{\alpha\beta\lambda} \delta^{3}(\xi -\eta)
$$
$$=\frac{1}{4}\int d^{3}zd^{3}z'[\partial_{\mu}A_{\nu}(z;x,y)-(
\mu \leftrightarrow \nu)]\delta^{3}(z-z')[\partial^{\mu}
A^{\nu}(z';x,y)-(\mu \leftrightarrow \nu )]
$$
$$
=\frac{1}{4} \int d^{3}z A^{2}_{\mu\nu}(z;x,y)=S_R[A_{\mu}]
\eqno(2.34)
$$

We immediately conclude that this term is canceled by the surface
renormalization counterterm in (2.33).

The second term in the exponent in (2.33) can be evaluated by making use
of the results of Appendix C.  It is surface independent and, in the
limit $\rho\rightarrow 0$ only depends on the points $x$ and $y$.
Indeed, according to (C.2) and already using (2.34)
$$
<\mu(x)\mu^*(y)>=\lim_{\epsilon\rightarrow 0}\exp\{-\frac{b^2}{2}
\sum^2_{i,j=1}\epsilon_i\epsilon_j[\frac{\pi}{|x_i-x_j|^2+|\epsilon|^2]
^{1/2}}]\}\eqno(2.35a)
$$
or
$$
<\mu(x)\mu^*(y)>=\lim_{\epsilon\rightarrow 0}\exp\{\pi b^2[\frac{1}
{|x-y|}-\frac{1}{\epsilon}]\}\eqno(2.35b)
$$
We now see explicitly that the renormalization term (2.7) indeed,
renders the $\mu$- correlation function surface independent.

The ultraviolet singularity in (2.35) may be eliminated by a
renormalization of the disorder(vortex) operator:$$
\mu_R(x)\equiv\lim_{\epsilon\rightarrow 0}\, \mu(x) \,
\exp[\frac{\pi b^2}{2|\epsilon|}]\eqno(2.36)$$

We therefore finally arrive at the result
$$
<\mu_R(x)\mu_R^*(y)>=\exp\{\frac{\pi b^2}{|x-y|}\}\eqno(2.37)
$$
for the disorder(vortex) euclidean correlation function in the Maxwell
theory.  Observe that
$$
\lim_{|x-y|\rightarrow \infty}<\mu_R(x)\mu^*_R(y)>=1\eqno(2.38)
$$
This indicates that $\mu_R$ does not create genuine vortex excitations,
in the true Maxwell theory.  One should expect this result since already
at the classical level the theory does not possess topological
excitations.  Another way of describing this fact is to say that there
is a vortex condensate in the Maxwell theory.  In section 3 we will
examine how this can be changed in the Maxwell-Higgs theory.

\bigskip
\leftline{ \large\bf 2.4) Order Correlation Function}

Let us evaluate now the two-point euclidean correlation function of
$\sigma$.  From(2.11) we see that there is no change in $\sigma$ as we
go to euclidean space.  Using the results of the previous subsection for
the euclidean action $S$ and gauge fixing term $S_{GF}$, as well as the
expression of $\sigma(x)$ eq.(2.11), we may write
$$
<\sigma(x)\sigma^*(y)>=\lim_{\rho ,\delta\rightarrow 0}Z^{-1}
\int DW_\mu\exp\{-\int d^3z[\frac{1}{2}W_\mu[P^{\mu\nu}+G^{\mu\nu}
]W_\nu+C_\mu(z;x,y)W^\mu]\}\eqno(2.39)
$$

In this expression, $C_\mu(z;x,y)=[C_\mu(z;x;C;T_x)-C_\mu(z;y;C;T_y)]$(as
in the case of $\mu$, the minus sign corresponds to the fact that we
have the adjoint operator$\sigma^*(y)$) where
$$C^\mu(z;x;C;T_x)=$$ $$
=\left \{\begin{array}{ll}
            ia\int _{R_x(C)} d^2\xi [\epsilon^{ki}\partial^{(\xi)}
            _k arg(\vec \xi -\vec x) +\partial^{(\xi)}_i\ln
            |\vec \xi -\vec x|] \delta^3(z-\xi) &\mbox{for $\mu =i$}\\
            0                                   &\mbox{for $\mu =0$}
         \end{array}
 \right.
\eqno(2.40)
$$

Performing the functional integral in (2.39), we get
$$
<\sigma(x)\sigma^*(y)>=\lim_{\rho ,\delta \rightarrow 0}\exp \{
\frac{1}{2} \int d^3zd^3z' C_\mu(z;x,y)
C_\nu(z';x,y)[P^{\mu\nu}+G^{\mu\nu}]^{-1}(z-z')\}\eqno(2.41)
$$

The $\sigma$ correlation function, of course is not gauge invariant
since $\sigma$ itself is not.  In order to obtain the gauge independent
part of $<\sigma\sigma^*>$, let us use in (2.41) just the gauge
independent part of $<W_\mu W_\nu>$, namely, the $\delta^{\mu\nu}$-
proportional part of $<W_\mu W_\nu>$ in (2.27b), which is explicitly
given by (2.32) (times $\delta^{\mu\nu}$).  Inserting this and (2.40)
in (2.41), and integrating on $z$ and $z'$, we get
$$
<\sigma(x)\sigma^*(y)>=\lim_{\rho ,\delta ,\epsilon\rightarrow 0}
\exp\{-\frac{a^2}{2}\sum^2_{i,j=1}\epsilon_i\epsilon_j
\int_{R_{x_i}(C)} d^2\xi \int _{R_{x_j}(C)} d^2\eta\times$$
$$
[\epsilon^{ki}\partial^{(\xi)}_k arg(\vec \xi -\vec x _i)+\partial^i
_{(\xi)}\ln |\vec \xi -\vec x _i|][\epsilon^{lj}\partial^{(\eta)}_l
arg(\vec \eta -\vec x _j) +\partial^j_{(\eta)}\ln |\vec \eta -\vec x
 _j|][\frac{\delta^{ij}}{4\pi[|\xi -\eta|^2 + |\epsilon |^2]^{1/2}}
]\}\eqno(2.42)
$$

Let us now use the identity
$$
\delta^{\mu\nu}\ \ \frac{e^{-m[|x|^2+|\epsilon |^2]^{1/2}}}{[|x|^2+|\epsilon
|^2]^{1/2}} = \partial^\mu \partial^\nu [-\frac{e^{-m[|x|^2+|\epsilon |
^2]^{1/2}}}{m}]+$$ $$+\frac{x^\mu x^\nu}{[|x|^2+|\epsilon |^2]}\ \ \
[\frac{1}{[|x|^2+|\epsilon |^2]^{1/2}}+m]\ \ \  e^{-m[|x|^2+|\epsilon |^2]
^{1/2}}\eqno(2.43a)
$$
which in the limit $m\rightarrow 0$ reduces to
$$
\frac{\delta^{\mu\nu}}{[|x|^2+|\epsilon |^2]^{1/2}} = \partial^\mu
\partial^\nu [-\frac{1}{m} +[|x|^2+|\epsilon |^2]^{1/2}]
+\frac{x^\mu x^\nu}{[|x|^2+|\epsilon |^2]^{3/2}}\eqno(2.43b)
$$
Insertion of (2.43b) in (2.42) produces two pieces.  The one coming from
the second term in (2.43b) vanishes in the limit
$\rho,\delta\rightarrow 0$ as is shown in Appendix C.  The one coming
from the first term in (2.43b) can be computed with the help of (C.1),
after using the fact that $\partial^\nu_{(\xi)}F(\xi
-\eta)=-\partial^\nu_{(\eta)} F(\xi-\eta)$.  We therefore conclude,
according to (C.1) that
$$
<\sigma(x)\sigma^*(y)>=\lim_{m,\epsilon\rightarrow 0} \exp \{
\frac{a^2}{8\pi}\sum^2_{i,j=1}\epsilon_i\epsilon_j(2\pi)^2
[-\frac{1}{m}+[|x_i-x_j|^2+|\epsilon|^2]^{1/2}]\}\eqno(2.44a)
$$
or
$$
<\sigma(x)\sigma^*(y)>=\exp\{-\pi a^2|x-y|\}
\eqno(2.44b)
$$

Observe that the infrared singularities at $m\rightarrow 0$ completely
cancel.  If we were calculating a charge nonconserving correlation
function (as $<\sigma\sigma>$, for instance) we would have
$\epsilon_i=\epsilon_j$ and the ${1\over m}$ factors would no longer
cancel, implying $<\sigma\sigma>=0$ as $m\rightarrow 0$.  We see that,
as in $1+1D$ [4], an infrared singularity is responsible for enforcing
the selection rule for the dual operators.  Observe that
$$
\lim_{|x-y|\rightarrow 0} <\sigma(x)\sigma^*(y)>=0\eqno(2.45)
$$
indicating that the charged states created by $\sigma$ are orthogonal to
the vacuum and charge is a conserved quantity as one would expect in the
pure Maxwell Theory.

\bigskip
\leftline{ \large\bf 2.5) The Mixed and Anyon Correlation Functions}
\bigskip

Let us examine now the mixed order-disorder euclidean correlation
function from which we will be able to obtain the anyon correlation
function.

Combining (2.26) with (2.39), we can write
$$
<\sigma(x_1)\mu(x_2)\sigma^*(y_1)\mu^*(y_2)>=\lim_{\rho ,\delta
\rightarrow 0} Z^{-1}\int DW_\mu \exp \{-\int d^3z[\frac{1}{2}
W_\mu[P^{\mu\nu}+G^{\mu\nu}]W_\nu+ $$
$$C_\mu (z;x_1,y_1)W^\mu + W_\mu P^{\mu\nu} A_\nu(z;x_2,y_2)
+\frac{1}{4} A_{\mu\nu}^2(z;x_2,y_2)]\}\eqno(2.46)
$$
The functional integral can be evaluated as before, yielding
$$
<\sigma(x_1)\mu(x_2)\sigma(y_1)\mu(y_2)>=\lim_{\rho ,\delta\rightarrow
0} \exp\{\frac{1}{2}\int d^3zd^3z'[B_\mu(z;x_2,y_2)+C_\mu(z;x_1,y_1)]$$
 $$\times
[B_\nu(z';x_2,y_2)+C_\nu(z';x_1,y_1)][P^{\mu\nu}+G^{\mu\nu}]^{-1}
(z-z')-S_R[A_\mu (x_2,y_2)]\}\eqno(2.47)
$$

The BPB and CPC terms were evaluated before.  The novel terms which
appear here are the BPC terms(already considering the two of
them):
$$
BPC=\int
d^3zd^3z'B_\mu(z;x_2,y_2)C_\nu(z';x_1,y_1)[P^{\mu\nu}+G^{\mu\nu}]^{-1}
(z-z')=\lim_{\rho ,\delta ,m\rightarrow 0} iab \sum^2_{i,j=1}
\epsilon_i\epsilon_j $$
$$\times
\int _{T_{r_i}(C)} d^2\xi_\mu \int
_{R_{s_j}(C)} d^2\eta \partial^{(\xi)}_\nu arg(\vec \xi -\vec r _i)
[\epsilon^{ki} \partial^{(\eta)}_k arg(\vec \eta -\vec s _j)  +
\partial^i \ln |\vec \eta - \vec s  _j|]$$
$$\times
F^{\mu\nu}_{(\xi)}\ _\sigma [P^{\sigma i}_{(\xi)} -\frac{1}{\xi}
\partial^\sigma_{(\xi)}\partial^i_{(\xi)}][\frac{1}{m}-
\frac{|\xi -\eta|}{8\pi}]\qquad\qquad (i,k=1,2)\eqno(2.48)
$$
Here $r_1\equiv x_2$, $r_2\equiv y_2$, $s_1\equiv x_1$ and
$s_2\equiv y_1$.  As before $\epsilon_1\equiv 1$ and $\epsilon_2\equiv
-1$.  As in the case of $<\mu\mu^*>$, the gauge dependent part of the
propagator in (2.48) gives no contribution, because $F^{\mu\nu}\ _\sigma
\partial^\sigma\equiv 0$. Only the $-\Box\delta^{\sigma i}$ part
contributes:
$$
F^{\mu\nu}\ _\sigma P^{\sigma i}=-\Box(\partial^\mu \delta^{\nu i}
-\partial^\nu\delta^{\mu i}){\cal F}^{-1}[\frac{1}{k^4}]=
(\partial^\mu \delta^{\nu i}- \partial^\nu\delta^{\mu i})
{\cal F}^{-1}[\frac{1}{k^2}]\eqno(2.49)
$$

The $\delta^{\mu i}$-part of (2.49) vanishes because $d^2\xi^\mu$ is
orthogonal to $R_{s_j(C)}$.  The remaining part gives
$$
BPC=\lim_{\rho ,\delta\rightarrow 0} iab
\sum^2_{i,j=1}\epsilon_i\epsilon_j \int _{R_{s_j}(C)} [\epsilon^{ki}
\partial^{(\eta)}_k arg(\vec \eta -\vec s_j)+ \partial_{(\eta)}^i
\ln|\vec \eta -\vec s _j|]  $$
 $$\times
\int _{T_{r_i}(C)} d^2\xi_\mu \partial_i^{(\xi)} arg(\vec \xi -\vec r
_i)\  \partial^\mu_{(\xi)}[\frac{1}{4\pi|\xi -\eta|}]\eqno(2.50)
$$
The above integrals are evaluated in Appendix D. Using (D.1), we get
$$
BPC=i\pi ab \sum^2_{i,j-1}\epsilon_i\epsilon_j arg(\vec s_j -\vec r_i)
$$
or
$$
BPC=i\pi ab[arg(\vec x_1-\vec x_2)+arg(\vec y_1-\vec y_2)
-arg(\vec x_1-\vec y_2)-arg(\vec y_1-\vec x_2)]\eqno(2.51)
$$
Combining (2.51) with the previous results for the BPB and CPC terms,
we obtain
$$
<\sigma(x_1)\mu_R(x_2)\sigma^*(y_1)\mu_R^*(y_2)>=\exp\{-\pi a^2
|x_1-y_1| + \frac{\pi b^2}{|x_2-y_2|}$$ $$+i\pi ab
[arg(\vec x _1-\vec x_2)+ arg(\vec y_1-\vec y_2)-arg(\vec x_1-\vec y_2)
-arg(\vec y_1-\vec x_2)]\}\eqno(2.52)
$$

The anyon correlation function can now be obtained in a straightforward
manner. Introducing the anyon field
$$
\varphi (x)=\lim_{x_1\rightarrow x_2\equiv x}\sigma(x_1)\mu_R(x_2)
\exp[-i\pi ab\ \  arg(\vec x_1-\vec x_2)]\eqno(2.53)
$$
we immediately get
$$
<\varphi(x)\varphi^*(y)>=\exp\{-\pi a^2|x-y|+\frac{\pi b^2}{|x-y|}
-i\pi ab[arg(\vec x-\vec y)+arg(\vec y-\vec x)]\}\eqno(2.54)
$$
Observe that $<\varphi\varphi^*>$ is multivalued, the ambiguous phase being
$\exp[i2\pi(\pi ab)]$.  As in the case of $1+1D$ [4],we interpret this fact
as being the manifestation in the framework of euclidean correlation
functions of the nontrivial commutation rule of $\varphi$.  Of course, the
same functional integral describes the two possible orderings of
operators in the left hand side of (2.54).  Making the analytic
continuation from euclidean to Minkowski space from each sheet of (2.54)
would reproduce each possible ordering of $\varphi$[4].  This, of course
implies that the spin of $\varphi$ is $s_\varphi = \pi ab$, confirming the
result we found above by direct computation of the $\varphi$ commutator.

The $\varphi$-correlation function (2.54) decays, at large distances as
$\exp[-\pi a^2 \vert x- y\vert]$. This implies the anyon states created
by $\varphi$ possess a mass $M=\pi a^2$.

\bigskip
\leftline{ \Large\bf 3) The Maxwell-Higgs Theory}
\vskip 2pt
 \leftline{\Large\bf (Abelian Higgs Model)}
\bigskip
\leftline{ \large\bf 3.1) Introduction}

Let us study in this section the properties of the dual operators
$\sigma$ and $\mu$ introduced before, in the framework of the Abelian
Higgs Model (AHM), described by
$$
{\cal L}=-\frac{1}{4} W_{\mu\nu}^2+ |D_\mu \phi |^2- g^2 \phi ^*\phi
-\frac{\lambda}{4}(\phi^*\phi )^2 \eqno(3.1)
$$
where $D_\mu =\partial _\mu +
ie W_\mu$.

The theory exists in two phases: i) a symmetric phase, for $g^2>0$,
where $<\phi>= 0$ and $W_\mu$ is massless and ii) a broken phase, for
$g^2<0$, where $<\phi>=\kappa\not= 0$ and $W_\mu$ acquires a mass $M=e\kappa$
through the Higgs mechanism.

Let us introduce in this theory the $\sigma$ and $\mu$ operators defined
in section 2.  Since the equal time commutators of the AHM are exactly
the same as in (2.12), we conclude that all commutation rules (equal
times) evaluated in section 2.2 for the pure Maxwell theory remain valid
here.

As the surface independence of $\mu$ is concerned, let us note that the
same counterterm (2.7) will be sufficient to make the $\mu$-correlation
functions surface independent here, because also in the AHM it will make
the integrand in the functional integral defining $<\mu\mu^*>$ to depend
on the external field through the combination $W_\mu +A_\mu$.  The
remaining three last terms in (3.1) will be invariant under the change
of variable (involving $\omega$, eq. (2.9)) needed to show surface
independence.  As was shown in [14], $\mu$ is the operator which creates
the quantum states associated to the classical soliton (vortex) solution
of Nielsen and Olesen [22].

In this interacting theory, of course, we will no longer be able to
compute exact correlation functions.  Instead, we will evaluate the long
distance behavior of them.  As we will see, this will be enough to
obtain interesting physical consequences.

\bigskip
\leftline{ \large\bf 3.2) The Unbroken Phase}

Let us study here the symmetric phase  $(g^2>0)$, in which $<\phi>=0$.
We start with the $\mu$ correlation function which is given by an
expression similar to (2.26):
$$
<\mu(x)\mu^*(y)>=\lim_{\rho , \delta\rightarrow 0}Z^{-1}
\int DW_\mu \exp \{-\int d^3z[\frac{1}{2}W_\mu[P^{\mu\nu}+G^{\mu \nu}]
W_\nu+ |D_\mu \phi |^2 + V(\phi )+$$
 $$
W_\mu P^{\mu\nu}A_\nu(z;x,y)+ \frac{1}{4}
A^2_{\mu\nu}(z;x,y)]\}\eqno(3.2)
$$

Using (2.28), we see that $<\mu\mu^*>$ is obtained by coupling the
external field $B_\mu (z;x,y)$ to the AHM lagrangian, in the way given
by (2.28a).  We therefore conclude that
$$
<\mu(x)\mu^*(y)>= \exp\{\Lambda(x,y) - S_R[A_\mu(x,y)]\}\eqno(3.3)
$$
where, using a diagrammatic language, $\Lambda(x,y)$ is given by the sum
of all Feynman diagrams with the field $B_\mu (z;x,y)$ in the external legs.

We are only interested here in the long distance behavior of
$<\mu\mu^*>$.  As was shown in [21], only two-leg graphs contribute to
(3.3) in this limit.  We are going to evaluate $\Lambda(x,y)$ by making a
loop expansion (in powers of $\hbar$).  The only vertex involving the
external field $B_\mu$ is given in Fig. 4a.  The lowest order (two-leg)
graph contributing to $\Lambda(x,y)$ is given in Fig. 4b.  Inserting the
euclidean propagator for $W_\mu$, eq. (2.27) we immediately see that the
graph of Fig. 4b. is identical to the first term in the exponent in
(2.29).  It follows that in this order of approximation $(0({1\over
\hbar})$, the large distance behavior of $<\mu\mu^*>$ is given exactly
by the same expression as in the Maxwell theory, namely
$$
<\mu_R(x)\mu^*_R(y)> \stackrel{|x-y|\rightarrow\infty}{\longrightarrow}
\exp\{\frac{\pi b^2}{|x-y|}\} \stackrel{|x-y|\rightarrow\infty}{
\longrightarrow} 1  \eqno(3.4)
$$

We have renormalized $\mu$ in the same way as in (2.36).

{}From (3.4) we conclude that in the symmetric phase of the AHM,  in
analogous way as in the true Maxwell theory, the disorder operator $\mu$
does not create states orthogonal to the vacuum, that is, genuine
excitations, as one should expect.  Also the symmetric phase of the AHM
can be viewed as a vortex condensate.

Let us turn now to the order correlation function. This is now given by
 $$
\cors =\lim_{\rho ,\delta\rightarrow 0} Z^{-1}\int DW_\mu
\exp\{ -\int d^3z[\frac{1}{2} W_\mu[P^{\mu\nu} + G^{\mu\nu}]W_\nu+$$
$$+
|D_\mu \phi |^2 + V(\phi )+ C_\mu(z;x,y)W^\mu ]\}\eqno(3.5)$$
where $C_\mu(z;x,y)$ is defined in (2.39) and (2.40).
Now,
$$
\cors =\exp\{\Gamma (x,y)\}\eqno(3.6)
$$
where $\Gamma(x,y)$ is the sum of all Feynman graphs containing the
field $C_\mu(z;x,y)$ in the external legs.  Again, only two-leg graphs
contribute to the long distance behavior of $<\sigma\sigma^*>$ [21].
The only vertex involving $C_\mu$ is the same as for $B_\mu$ and is
shown in Fig. 4a.  Again, making an expansion in loops, the lowest order
graph will be the one of Fig. 4b.  Inserting the $W_\mu$-propagator,
eq.(2.27), we conclude that in this order ($O(\frac{1}{\hbar})$)
the large distance behavior of
the gauge invariant part of $<\sigma\sigma^*>$ is given by the same
expression as in the Maxwell theory, namely
$$
\cors \stackrel{|x-y|\rightarrow\infty}{\longrightarrow}
\exp\{-\pi a^2|x-y|\}\stackrel{|x-y|\rightarrow\infty}{\longrightarrow}
 0\eqno(3.7)
$$

The long distance behavior of the anyon field could be obtained as well,
by exchanging $B_\mu(z;x,y)$ for $B_\mu(z;x,y) + C_\mu(z;x,y)$ in (3.2).  In
the lowest order of approximation, we would get
$<\varphi(x)\varphi^*(y)>$ behaving asymptotically as (2.54).

\bigskip
\leftline{\large\bf 3.3) The Broken Phase}
\bigskip
\leftline{\bf 3.3.1) The Order Correlation Function}

Let us consider now the case in which $g^2<0$ and
$<\phi>=\kappa =({4|g^2|\over \lambda})^{1\over 2}$.  We are going to write
$\phi$ in terms of its real components: $\phi = {1\over\sqrt{2}}(\phi_1
+i \phi_2)$ and choose $<\phi_1>=\kappa$ and $<\phi_2>=0$.

Let us compute here the order correlation function$<\sigma\sigma^*>$.
This is given by (3.5), and (3.6). Again, in lowest order
$<\sigma\sigma^*>$ will be given by the graph of Fig. 4a.  There is,
however, an important difference.  We must shift the Higgs field around
the vacuum value.  This will generate the following quadractic part for
the euclidean lagrangian ${\cal L}[W_\mu,\phi_1,\phi_2]$
$$
{\cal L}^{(2)}[W_\mu ,\phi_1,\phi_2]=\frac{1}{4} W_{\mu\nu}^2
+\frac{1}{2}M^2W_\mu W_\mu +\frac{1}{2} (\partial_\mu \phi_1)^2+
\frac{1}{2}m_1^2\phi_1^2 + \frac{1}{2}(\partial_\mu \phi_2)^2 +
MW_\mu \partial^\mu \phi_2  \eqno(3.8)
$$
where $M=e\kappa$ and $m_1=2|g^2|$.

Using the Lorentz gauge $(\xi\rightarrow\infty)$, we see that $W_\mu$
and $\phi_2$ decouple in the quadractic part (since $\partial_\mu W^\mu
=0$).  The euclidean propagators  for $W_\mu$, $\phi_1$ and $\phi_2$
are now
given respectively by
$$
D_{\mu\nu}(k)=\lim_{\xi\rightarrow \infty} \
[\frac{P^{\mu\nu}}{k^2(k^2+M^2)}+
\frac{k^\mu k^\nu}{k^2(\xi k^2+M^2)}]\eqno(3.9a)
$$
$$
\Delta_1(k)=\frac{1}{k^2+m_1^2}\ \ \ \ \ ; \ \ \ \ \ \Delta_2(k)=\frac{1}
{k^2} \eqno(3.9b)
$$

Inserting the $W_\mu$-propagator in the graph of Fig. 4b, we see that,
the lowest order $(0({1\over \hbar}))$ contribution to large distance
behavior of the gauge invariant part of $<\sigma\sigma^*>$ is given by
an expression identical to (2.41) and (2.42), except for the fact that
the last term between square brackets in (2.42), namely
${\cal F}^{-1}[{1\over
k^2}]$, is exchanged by ${\cal F}^{-1}[{1\over k^2+M^2}]$, that is
$$
\delta^{\mu\nu}{\cal F}^{-1}[\frac{1}{k^2+M^2}]=\lim_{\epsilon
\rightarrow 0}\  \delta^{\mu\nu}\  \frac{e^{-[|x|^2+|\epsilon |^2]
^{1/2}}}{[|x|^2+|\epsilon |^2]^{1/2}} \eqno(3.10)
$$
Using the identity (2.43a) and following the same steps which led us to
(2.44), we immediately conclude that in lowest order $(0({1\over
\hbar}))$,
$$
\cors \stackrel{|x-y|\rightarrow\infty}{\longrightarrow}
\exp\{\frac{\pi a^2}{M}[e^{-M|x-y|}-1]\}\eqno(3.11a)
$$
or
$$
<\sigma_R(x)\sigma^*_R(y)>
\stackrel{|x-y|\rightarrow\infty}{\longrightarrow}
\exp\{\frac{\pi a^2}{M}e^{-M|x-y|}\}\stackrel{|x-y|\rightarrow\infty}
{\longrightarrow}   1 \eqno(3.11b)
$$
where $\sigma_R\equiv\sigma \exp[{\pi a^2\over 2 M}]$.

Observe that in the broken phase, the charge bearing operator $\sigma$
no longer creates states orthogonal to the vacuum.  This corresponds to
the screening of charge associated with the mass generation to the field
$W_\mu$ through the Higgs mechanism.  Notice that in the limit
$M\rightarrow 0$, $<\sigma\sigma^*>$ reduces to the expression found in
(3.7) for the symmetric phase and the zero mass singularities cancel in
charge selection rule respecting correlation function.

\bigskip
\leftline{\bf 3.3.2)The Disorder (Vortex) Correlation Function}

Let us evaluate here the $\mu$-correlation function in the broken phase
of the AHM.  As we infer from (3.2), we can write $<\mu\mu^*>$ as
$$
\corm =\lim_{\rho ,\delta\rightarrow 0} Z^{-1}\int DW_\mu\exp\{-S
[W_{\mu\nu} +A_{\mu\nu} , \phi ,D_\mu \phi ] -S_{GF}[W_\mu]\}
\eqno(3.12)
$$
where $S[W_{\mu\nu},\phi, D_\mu\phi]$ is the action associated with
(3.1).

Performing the change of variable $W_\mu\rightarrow W_\mu + A_\mu(z;x,y)$,
we get
$$
\corm =\limrd \int D_\mu \exp\{-S[W_{\mu\nu},\phi,\widetilde{D}_\mu \phi]
-S_{GF}[W_\mu-A_\mu]\}\eqno(3.13)
$$
where $\widetilde D_\mu =\partial_\mu+ ie[W_\mu-A_\mu(x,y)]$.  It will be more
convenient to use (3.13) to compute $<\mu\mu^*>$ in the broken phase.
Now, we can write
$$
\corm =\exp\{\tilde{\Lambda}(x,y)\}\eqno(3.14)
$$
where $\widetilde\Lambda(x,y)$ is the sum of all Feynman graphs with the
$A_\mu(z;x,y)$ field in the external legs and computed with the Feynman
rules coming from (3.13).  We are again going to make an expansion in
loops and furthermore, in powers of $M$, the gauge field mass.  Before
evaluating $\widetilde\Lambda(x,y)$, we shift the Higgs field around the vacuum
value in (3.13).  This operation commutes with the change of variable
$W_\mu\rightarrow W_\mu + A_\mu$.  The vertices relevant for the lowest
order computation of $\widetilde\Lambda(x,y)$ are depicted in Fig. 5.  Their
contribution, in this order,
 is given by the graphs of Fig. 6.  The gauge dependent terms,
which appear in
Fig. 6a, vanish for all values of $\xi$ as can be easily seen from
(3.9a).  The first nonzero contribution for $\widetilde\Lambda(x,y)$, therefore
is given by the graphs of Fig. 6b, which are of the order $0({M^2\over
\hbar})$.  Using (3.9), in the graphs of Fig. 6b, we find
$$
\tilde{\Lambda}(x,y)\tendei -\frac{M^2}{2}\int d^3zd^3z'A_\mu(z;x,y)
A_\nu(z';x,y)P^{\mu\nu}F(z-z') \eqno(3.15)
$$
where $F(z-z')={\cal F}^{-1}[{1\over k^2}]=\Delta_2(z-z')$ and is given by
(2.32).
Eq. (3.15) can be written as
$$
\tilde{\Lambda}(x,y) \tendei  \frac{M^2}{2}\int d^3zd^3z'\epsilon
^{\mu \alpha\lambda}\partial^{(z)}_\alpha A_\mu(z;x,y)F(z-z')
\epsilon^{\nu\beta\lambda}\partial^{(z')}_\beta A_\nu(z';x,y)
$$

Using the result of Appendix B, we get, after integrating over $z$ and
$z'$ and using (2.32):
$$
\tilde{\Lambda}\tendei \limrde -\frac{M^2b^2}{2} \soma \inte $$
$$\times
\epsilon^{\mu\nu\lambda}\epsilon^{\alpha\beta\gamma}
[\frac{\delta^{\lambda\gamma}}{4\pi [|\xi -\eta|^2+|\epsilon |^2]
^{1/2}}] \eqno(3.16)
$$

Making use of the identity (2.43b) we see that (3.16) contains two
terms, corresponding to the two pieces in the right hand side of (2.43b).
The second one is surface dependent but, as was shown in [17], for
$i\not= j$, it vanishes in the limit $|x-y|\rightarrow\infty$. This
therefore does not contribute to (3.16).  The $i=j$ term is just a
self-energy that will renormalize $\mu$ [17].  The first term of (3.16) can be
evaluated immediately with the help of result (C.2) of Appendix C:
$$
\tilde{\Lambda} \tendei \lim_{\epsilon\rightarrow 0}
\frac{\pi M^2b^2}{2} \soma [-\frac{1}{m} +[|x_i-x_j|^2+|\epsilon |
^2]^{1/2}] \eqno(3.17a)
$$
or
$$
\tilde{\Lambda}(x,y) \tendei -\pi M^2b^2|x-y| \eqno(3.17b)
$$

We therefore conclude that

$$
<\mu_R(x)\mu^*_R(y)> \tendei \exp\{-\pi M^2b^2|x-y|\} \tendei 0
\eqno(3.18)
$$

We see that in the broken phase the vortex operator $\mu$ indeed creates
states orthogonal to the vacuum, i.e., genuine excitations in the
spectrum.  According to (3.18), the mass of these excitations is $M_v
=\pi M^2b^2$, up to the order of approximation we are working
$(0({M^2\over \hbar}))$.  It is interesting to see that choosing $b^2=e^{-2}$
[21], the mass of the quantum vortices created by $\mu$ coincides with
the classical vortex energy found in [23].  In ref. [21] one-loop
corrections to the above result were also evaluated.

Our study of the AHM and Maxwell theory clearly exposes the reason why
we call $\sigma$ and $\mu$ ``order'' and ``disorder'' operators.
$\sigma$ has a nonzero vacuum expectation value in the broken (ordered)
phase, while the vacuum expectation value of $\mu$ is different from
zero in the symmetric (disordered) phases and vice-versa.  In the broken
(ordered) phase $\mu$ creates genuine quantum soliton states.  The
transition to the symmetric (disordered) phases can be viewed as a
condensation of topological charge in the same way as the transition to
a broken (symmetric) phase can be viewed as a condensation of charge as
occurs
in superconductivity.  We see that in a certain sense, the AHM is quite
similar to the Ising model, the phase transition which generates mass to
the gauge field being analogous to the phase transition which takes
place in the latter.

A final word is in order about the procedure adopted for the computation
of $<\mu\mu^*>$ in the broken phase of the AHM.  Observe that we shifted
$W_\mu$ around $A_\mu$ before evaluating this correlation function,
contrary to what we did in other cases.  We proceeded this way because
without shifting, the lowest order in our approximation scheme would
produce a trivial result as the computations would be effectively done
in the Proca theory.  We would have to go to higher orders in order to
get a nontrivial result, while with the shifting trick we already get it
in lowest order.  Of course the two ways of calculating should produce
the same result if one was able to sum the whole perturbation series.

\bigskip
\leftline{\Large\bf 4) The Maxwell-Chern-Simons Theory}
\bigskip
\leftline{\large\bf 4.1) Order and Disorder (Vortex) Operators}

Let us study now the Maxwell-Chern-Simons (MCS) theory, described by
$$
{\cal L}=-\frac{1}{4}W_{\mu\nu}W^{\mu\nu} +\frac{\theta}{2}
\epsilon^{\mu\nu\alpha}W_\mu\partial_\alpha W_\beta \eqno(4.1)
$$
which leads to the field equation
$$
\partial_\nu W^{\mu\nu}= \theta \epsilon^{\mu\alpha\beta}
\partial_\alpha W_\beta \eqno(4.2)
$$
The parameter $\theta$ has dimension of mass.

In addition to the identically conserved current (2.3), let us introduce
the two indexes current $\widetilde J ^{\mu\nu}_\theta$ which
generalizes (2.4):
$$
\tilde{J}^{\mu\nu}_\theta =W^{\mu\nu}- \theta \epsilon^{\mu\nu\alpha}
W_\alpha \eqno(4.3)
$$
$\widetilde J ^{\mu\nu}_\theta$ is conserved as a consequence of the
field equation (4.2) in analogy with (2.4).

Let us introduce also the disorder operator $\mu_\theta$ which generalizes
(2.6) for $\theta\not= 0$, by coupling $\widetilde J^{\mu\nu}_\theta$ to
$A_{\mu\nu}$:
$$
\mu_\theta(x;C)=\exp\{-\frac{ib}{2}\int d^3z \tilde{J}^{\mu\nu}_\theta
A_{\mu\nu}\} $$
or
$$
\mu_\theta(x;C)=\exp\{\frac{ib}{2}\int d^3z W^{\mu\nu}A_{\mu\nu}+
ib\theta \int d^3z \epsilon^{\mu\alpha \beta}W_\mu \partial_\alpha
A_\beta\}
\eqno(4.4)
$$

We see that $\mu_\theta\equiv\mu(b)\Sigma '(\theta b)$, where $\Sigma '$
is given by (2.10b).

Making use of the surface renormalization counterterm (2.7), we
immediately see that the integrand in the functional integral defining
correlation functions of $\mu_\theta$ will depend on $A_\mu$ through the
combination $W_\mu+A_\mu$ and therefore, we expect these correlation
functions to be surface independent as before.  We are going to see
explicitly that this is indeed the case.

The $\sigma$ operator needs no modification in the MCS theory.  It is
given, as before, by eq. (2.11).

\bigskip
\leftline{\large\bf 4.2) Commutation Rules}

Let us obtain here the relevant commutators involving the $\sigma$ and
$\mu_\theta$ operators in the MCS theory.

The momentum canonically conjugate to $W^i$ is now given by $\Pi^i=-E^i
+{\theta\over 2} \epsilon ^{ij} W^j$, with $E^i=W^{io}$ [1].  The basic
commutators of the MCS theory are [1] (equal times)
$$
[W^i,W^j]=[\Pi^i,\Pi^j]=0
$$
$$
[W^i(x),E^j(y)]=-[W^i(x),\Pi^j(y)]= -i\delta^{ij}(\vec x - \vec y)
\eqno(4.5)
$$
$$
[E^i(x),E^j(y)]=-i\theta \epsilon^{ij}\delta^2(\vec x-\vec y)
$$
Observe that contrary to the Maxwell case, the electric field $E^i$ no
longer commutes with itself.

Using the results of Appendix B, we see that $\mu_\theta$ can be written
as
$$
\mu_\theta(x;C)=\exp\{-ib\int _{T_x(C)} d^2\xi [E^i(x^0,\vec \xi) +
\theta \epsilon^{ij}W_j(x^0,\vec\xi)]\partial_i arg(\vec\xi-\vec x)\}
\eqno(4.6)
$$
Using (4.5) we immediately see that the commutation rule between
$\sigma$ and $\mu_\theta$ in the MCS theory is identical to that in the
Maxwell theory involving $\sigma$ and $\mu$, namely (2.16) or (2.17).
Also in the MCS theory, we have $[\sigma,\sigma]=0$.  $\mu_\theta$,
however, no longer commutes with itself.  Writing $\mu_\theta(x;C)\equiv
e^{A(x;C)}$, we have according to (4.5),
$$
[A(x;C),A(y;C)]=\limrd -i\theta b^2 \int _{T_x(C)} d^2\xi \int
_{T_y(C)} d^2\eta \partial^{(\xi)}_i arg(\vec\xi -x)
\partial^{(\eta)}_j arg(\vec\eta -\vec y)
$$
$$\times
\epsilon^{ij}
\delta^2(\xi -\eta) \eqno(4.7a)
$$
or
$$
[A(x;C),A(y;C)]= \limrd -i\theta b^2\int _{T_x(C)\cap T_y(C)}
d^2\xi \epsilon^{ij} \partial_iarg(\vec\xi-\vec x)
\partial_j arg(\vec\xi -\vec y) \eqno(4.7b)
$$
It is shown in Appendix C, that, after taking the limit
$\rho,\delta\rightarrow 0$, in $T_x(C)$ and $T_y(C)$, eq. (4.7b) is
given by
$$
[A(x;C),A(y;C)] \limrd -i2\pi \theta b^2\ [arg(\vec x -\vec y)-
arg(\vec y-\vec x)]$$
$$
\,\,\,\,\,\,\,\,\,\,\,\,\,\,=-i2\pi (\pi \theta b^2)\ \epsilon
[arg(\vec x -\vec y) - \pi ] \eqno(4.8)
$$
where $\epsilon(x)= sign(x)$.  It follows that, at equal times
$$
\mu_\theta(x)\mu_\theta(y)=\mu_\theta(y)\mu_\theta(x)
\exp\{-i2\pi (\pi \theta b^2)\epsilon[arg(\vec x-\vec
y)-\pi]\}\eqno(4.9)
$$
indicating that in the MCS theory the disorder (vortex) operator is
itself anyonic, carrying statistics $S_{\mu_\theta} =\pi\theta b^2$.

Let us investigate now whether the commutation rules of $\sigma$ and
$\mu_\theta$ with the charge and magnetic flux operators, (2.18), are
changed in the MCS theory.

According to (4.5), we immediately see that the commutation rules
involving $\sigma$ are the same as in Maxwell theory, namely, eq. (2.20)
remains valid, indicating that $\sigma$ bears charge but not magnetic
flux.  The commutator between $\mu_\theta$ and the magnetic flux
(topological charge) operator is also identical to the one in Maxwell
theory.  Let us evaluate the commutator between $\mu_\theta$ and the
charge operator.  Writing $\mu_\theta (x)\equiv e^{\alpha(x)}$ and
$\rho(y)=\partial_j E^j (y)\equiv\beta(y)$, we have, according to (4.5)
$$
[\alpha(x),\beta(y)]=\limrd -ib\int _{T_x(C)} d^2\xi\  \partial^{(\xi)}
_i arg(\vec \xi-\vec x) \partial^{(y)}_j [-i\theta\epsilon^{ij} +
i\theta\epsilon^{ik}\delta^{jk}]\delta^2(\xi -y)=0 \eqno(4.10)
$$
We see that also in MCS theory we have $[\mu_\theta,Q]=0$ and eq.
(2.22) remains fully valid.

A very interesting phenomenon occurs in the MCS theory, involving the
statistics of $\mu_\theta$.  In spite of bearing only magnetic flux
(and no net charge) $\mu_\theta$ has generalized statistics
$S_{\mu_\theta}=\pi\theta b^2$.  This can be understood as follows.  In
the pure Maxwell theory, the two quantities determining the statistics
were the magnetic flux and the charge, whose densities were given
respectively by the zeroth components of $-J^\mu$ and $\widetilde
J_\mu\equiv \partial_\nu\widetilde J^{\nu\mu}$, where $\widetilde
J^{\nu\mu}$ is given by (2.4).  In the MCS theory, however, $\widetilde
J^{\mu\nu}$ must be exchanged by $\widetilde J^{\mu\nu}_\theta$, eq.
(4.3).  It follows that in addition to the magnetic flux, the other
quantity relevant for determining the statistics in the MCS theory will
have its density given by the zeroth component of $\widetilde
J^\mu_\theta\equiv\partial_\nu \widetilde J^{\nu\mu}_\theta$, or
$\widetilde J^0_\theta =\rho-\theta B$.  In other words, the statistics
in the MCS theory will be determined by the product of $\Phi$ and the
effective charge $\widetilde Q= Q-\theta\Phi$.  Indeed, we see that
according to (2.22) (which is also valid for $\mu_\theta$) $\mu_\theta$
carries $2\pi b$ units of $\Phi$ and $-2\pi\theta b$ units of
$\widetilde Q$, implying that it will have statistics
$S_{\mu_\theta}=|{\widetilde Q\Phi\over 4\pi}| =\pi\theta b^2$ (Observe
also the sign in (4.9).

A composite anyon operator can also be constructed in the MCS theory
through an expression like (2.23) or (2.53). Using the
$\sigma-\mu_\theta$ and $\mu_\theta-\mu_\theta$ commutation rules, we
immediately find the equal-time commutator
$$
\varphi(x)\varphi(y)=\varphi(y)\varphi(x) \exp\{i2\pi [\pi ab -
\pi \theta b^2]\  \epsilon[arg(\vec x -\vec y) -\pi]\}\eqno(4.11)
$$
This indicates that the composite anyon field has now statistics
$S_\varphi =\pi b|a-\theta b|$.  The field $\varphi$ carries magnetic
flux $\Phi=2\pi b$ and effective charge $\widetilde Q= 2\pi(a-\theta b)$.  We
see that here also the formula $S_\varphi = {|\widetilde Q\Phi|\over
4\pi}$ holds true.  For the special case $a=\theta b$, $\varphi$ will be
a boson, corresponding to the fact that $\widetilde Q =0$ for this value
of \underline{a}.

\bigskip
\leftline{\large\bf 4.3) Order Correlation Functions}

Let us compute now the euclidean order correlation function in the MCS
theory. In order to do this we will need the analytic continuation of
the Chern-Simons action to euclidean space:
$$
iS_{CS}=\frac{i\theta}{2}\int d^3z \epsilon^{\mu\alpha\nu}W_\mu
\partial_\alpha W_\nu \rightarrow -\int d^3z_E\frac{-i\theta}{2}
\epsilon^{\mu\alpha\nu} W_\mu \partial_\alpha W_\nu$$
 $$ \equiv
-\frac{\theta}{2} \int d^3z_E W_\mu C^{\mu \nu}W_\nu\equiv
-S_{CS,E}\,\,\,\,\, ;\,\,\,C^{\mu\nu}\equiv-i\epsilon^{\mu\alpha\nu}
\partial_\alpha  \eqno(4.12)
$$
Following the same steps as in the case of Maxwell theory and using
(4.12), we arrive at an expression similar to (2.39) for
$<\sigma\sigma^*>$:
$$
\cors =\limrd Z^{-1}\int DW_\mu \exp\{-\int d^3z[\frac{1}{2} W_\mu
[P^{\mu\nu}+\theta C^{\mu\nu}+G^{\mu\nu}]W_\nu+
$$
$$
C_\mu(z;x,y)W^\mu ]\}\eqno(4.13)
$$
where $C_\mu(z;x,y)$ was defined in (2.39-40).

Integrating over $W_\mu$ we readily obtain
$$
\cors =\limrd \exp\{\frac{1}{2}\int d^3z d^3z' C_\mu(z;x,y)C_\nu(z';x,y)
[P^{\mu\nu}+\theta C^{\mu\nu}+G^{\mu\nu}]^{-1}(z-z')\} \eqno(4.14)
$$
where $[P^{\mu\nu} + \phi C^{\mu\nu} + G^{\mu\nu}]^{-1}\equiv <W^\mu
W^\nu>_{MCS}$ is the euclidean propagator of the $W_\mu$ field in the
MCS theory. This is given in momentum space by
$$
<W^\mu(k)W^\nu(-k)>_{MCS}=[P^{\mu\nu}(k)-\theta\epsilon^{\mu\alpha\nu}
k_\alpha][\frac{1}{k^2(k^2+\theta ^2)}]+\frac{k^\mu k^\nu}{\xi k^4}
\eqno(4.15a)
$$

and in coordinate space by
$$
<W^\mu(x)W^\nu(y)>_{MCS}=[P^{\mu\nu}+i\theta\epsilon^{\mu\alpha\nu}
\partial_\alpha][\frac{1-e^{-\theta|x-y|}}{4\pi\theta^2|x-y|}]
-\lim_{m\rightarrow 0}\frac{1}{\xi}\partial^\mu\partial^\nu
[\frac{1}{m}-\frac{|x-y|}{8\pi}]\eqno(4.15b)
$$
Here, the second expression between brackets is ${\cal F}^{-1}[{1\over
k^2(k^2+\theta^2)}]$ and the third is ${\cal F}^{-1}[{1\over k^4}]$ (as
before, $m$ is an infrared regulator).  Observe that (4.15) reduces to
(2.27) in the limit $\theta\rightarrow 0(m\equiv {1\over 4\pi\theta}$ in
this limit).

As before, we want to extract the gauge independent part of
$<\sigma\sigma^*>$.  This will be achieved by inserting the gauge
independent part of $<W^\mu W^\nu>_{MCS}$, namely the $\delta^{\mu\nu}$
and $\epsilon^{\mu\alpha\nu}$- proportional terms of (4.15), in (4.14).
The $\delta^{\mu\nu}$- proportional part of (4.15) is
$$
-\Box[\frac{1-e^{-\theta |x-y|}}{4\pi\theta^2|x-y|}]= -\Box{\cal F}
^{-1}[\frac{1}{k^2(k^2+\theta^2)}]={\cal F}^{-1}[\frac{1}{k^2+\theta^2}]
=\frac{e^{-\theta |x-y|}}{4\pi |x-y|}\eqno(4.16)
$$
We can now use the identity (2.43a) and go through the same steps which
led to (2.44).  As before, the second term in (2.43a) will give a null
contribution in the limit $\rho,\delta\rightarrow 0$, as is shown in
Appendix C.  The $\epsilon^{\mu\alpha\nu}$-proportional term in (4.15)
also gives a vanishing contribution in this limit, for the same reason
(see Appendix C).

The contribution from the first term in (2.43a) can be evaluated in the
same way as we did in (2.44), leading to
$$
\cors =\lim_{\epsilon\rightarrow 0}\exp\{\frac{a^2}{8\pi} \soma
(2\pi )^2[\frac{\exp\{-\theta[|x_i-x_j|^2+|\epsilon |^2]\}}{\theta}]\}
\eqno(4.17a)
$$
$$
\cors =\exp\{\frac{\pi a^2}{\theta}[e^{-\theta |x-y|}-1]\}\eqno(4.17b)
$$
or
$$<\sigma_R(x)\sigma^*_R(y)>=\exp\{\frac{\pi a^2}{\theta}e^{-\theta
|x-y|}\} \eqno(4.17c)
$$
where $\sigma_R\equiv \sigma\exp [{\pi a^2\over 2\theta}]$.  Observe that
(4.17) reduces to (2.44) in the limit $\theta\rightarrow 0$.

Now, $$
\lim_{|x-y|\rightarrow \infty}<\sigma_R(x)\sigma^*_R(y)>=1\eqno(4.18)
$$
This result reflects the screening of charge associated with the mass
generation to the gauge field induced by the Chern-Simons term.

\bigskip
\leftline{\large\bf 4.4) Disorder (Vortex) Correlation Function}

Let us evaluate here the $\mu_\theta$- correlation function in the MCS
theory.  Using (4.4), (4.12), (2.25) and (2.7), we can write
$$
<\mu_\theta(x)\mu_\theta^*(y)>=\limrd Z^{-1} \int DW_\mu \exp\{
-\int d^3z[\frac{1}{2}W_\mu[P^{\mu\nu}+\theta C^{\mu\nu} +G^{\mu\nu}]
W_\nu + $$
$$
W_\mu[P^{\mu\nu}+\theta C^{\mu\nu}]A_\nu(z;x,y) +\frac{1}{4}
A^2_{\mu\nu}(z;x,y)]\}\eqno(4.19)
$$

 where $A_\mu(z;x,y)$ was defined in (2.26).

Notice that, since $S_{CS}[A_\mu]\equiv 0$ because $A_\mu$ only has the
$\mu =3$ component different from zero, the integrand in (4.9) depends
on $A_\mu$ through the combination $W_\mu+A_\mu$.  The method employed
in [14], therefore, can also be used to prove the surface invariance of
$<\mu_\theta\mu^*_\theta>$ in the MCS theory.

Integrating over $W_\mu$ with the help of (4.15) and using the results
of Appendix B, we get
$$
<\mu_\theta(x)\mu_\theta^*(y)>=\limrd \exp\{\frac{1}{2} \int d^3zd^3z'
[B_\mu(z;x,y)+D_\mu(z;x,y)]$$
 $$\times [B_\nu(z';x,y)+D_\nu(z';x,y)]
[P^{\mu\nu}+\theta C^{\mu\nu}+ G^{\mu\nu}]^{-1}(z-z')-S_R[A_\mu]\}
\eqno(4.20)
$$
In the above expression, $B_\mu(z;x,y)$ was defined in (2.28) and (2.29)
and according to (B.5), $D_\mu(z;x,y)=D_\mu(z;x)-D_\mu(z;y)$, where
$$
D^\mu(z;x)= -i\theta b\int_{T_x(C)} d^2\xi_\alpha
\epsilon^{\alpha\beta\mu}\partial_\beta arg(\vec \xi -\vec x)\eqno(4.21)
$$
Inserting $B_\mu,D_\mu$ and (4.15) in (4.20), we will get six terms:
BPB, BPD and DPD, corresponding to the $[P^{\mu\nu}
+\partial^\mu\partial^\nu]$ part of (4.15) and BCB, BCD and DCD,
corresponding to the $\epsilon^{\mu\alpha\nu}$ part of (4.15).  One of
these, namely DPD will be gauge dependent, as a consequence of the fact
that the $\theta$-dependent part of $\mu_\theta$ is not gauge invariant.  As
before, we will only consider the gauge independent part of this term.
Let us evaluate now each one of the six terms above.

The BPB term can be computed exactly as in (2.30), except for the fact
that now the $P^{\sigma\lambda}$-proportional term between square
brackets in (2.30) is replaced by ${\cal F}^{-1}[{1\over k^2(k^2+\theta^2)}]$
as
one can infer from (4.15).  We will arrive at the expression in the exponent
of (2.33) but with (2.32) replaced by (4.16) in the square brackets in
(2.33).  Following the same steps that we took after (2.33) we arrive at
$$
BPB=\lim_{\epsilon\rightarrow 0} \pi b^2[\frac{e^{-\theta |x-y|}}{|x-y|}
-\frac{1}{|\epsilon |}]+\frac{1}{4} \int d^3zd^3z' A_{\mu\nu}(z;x,y)F(z-z')
A_{\mu\nu}(z';x,y)\eqno(4.22)
$$
where
$$
F(z-z')={\cal F}^{-1}[\frac{k^2}{k^2+\theta ^2}]=-\Box \frac{
e^{-\theta |z-z'|}}{4\pi |z-z'|} \eqno(4.23)
$$
Observe that the last term in (4.22) is surface dependent but no longer
canceled by the renormalization counterterm $S_R[A_\mu]$.

Let us obtain now the BCD term.  This is given by( notice the factor two
because there are actually two crossed terms)
$$
BCD=\limrd \theta^2b^2 \soma \inte
$$
$$\times
 F^{\mu\nu}_{(\xi)}\ _\sigma
\epsilon^{\alpha\beta\gamma}\epsilon^{\sigma\rho\lambda}
\partial^{(\xi)}_\rho [\frac{1-e^{-\theta |x-y|}}{4\pi\theta^2 |\xi
-\eta|}\eqno(4.24)
$$
 Using an
identity similar to (2.31), namely,
$$
F^{\mu\nu}\ _\sigma \epsilon^{\alpha\beta\lambda}\epsilon^{\sigma\rho
\lambda}\partial_\rho=\epsilon^{\mu\nu\lambda}\epsilon^{\alpha\beta\gamma}
(-\Box
\delta^{\lambda\gamma}+\partial^\lambda\partial^\gamma)\eqno(4.25)
$$

we arrive at an expression identical to the exponent in (2.33) except
for the prefactor and for the expression between square brackets which
is replaced by ${\cal F}^{-1}[{1\over k^2(k^2+\theta^2)}]$.  Following the same
steps we took after (2.33) we immediately obtain
$$
BCD=2\pi\theta^2 b^2[\frac{1-e^{-\theta |x-y|}}{\theta |x-y|}-1]+
\frac{2\theta^2}{4}\int d^3zd^3z' A_{\mu\nu}(z;x,y)G(z-z')A_{\mu\nu}
(z';x,y) \eqno(4.26)
$$
  where
$G(z-z')={\cal F}^{-1}[{1\over k^2+\theta^2}]$ is given by the last equality in
(4.16).  Again, the last term in (4.26) is surface dependent.

Let us now consider the DPD term.  As we remarked before, it is gauge
dependent.  In order to get the gauge invariant part, let us take the
$\delta^{\mu\nu}$-proportional part of (4.15).  Then, using (4.21), we
get an expression identical to (2.34), except for the prefactor and for
the $\delta^3(\xi-\eta)$ function which is replaced by ${\cal F}^{-1}[{1\over
k^2+\theta^2}] =G(z-z')$, namely
$$
DPD=-\frac{\theta^2}{4}\int d^3zd^3z'A_{\mu\nu}(z;x,y)G(z-z')A_{\mu\nu}
(z';x,y)\eqno(4.27)
$$
The gauge invariant part of DPD only contains the surface dependent term
above.

Let us evaluate now the crossed BPD term.  This is given by
$$
BPD=\lim_{\rho ,\delta ,m\rightarrow 0}-i\theta b^2 \soma \inte $$
$$\times
F^{\mu\nu}_{(\xi)}\ _\sigma \epsilon^{\alpha\beta\lambda}
\{P^{\sigma\lambda}_{(\xi)} [\frac{1-e^{-\theta |\xi
-\eta|}}{4\pi\theta^2|\xi -\eta|}] -\frac{1}{\xi}\partial^\sigma
\partial^\lambda[\frac{1}{m}-\frac{|\xi -\eta |}{8\pi} ]\}\eqno(4.28)
$$

All the gauge dependent (derivative) terms vanish because $F^{\mu\nu} \
_\sigma\partial^\sigma\equiv 0$.  Let us observe now that
$$
F^{\mu\nu}_{(\xi)}\ _\sigma
\epsilon^{\alpha\beta\lambda}P^{\sigma\lambda}_{(\xi)}=-\Box
(\partial^\mu_{(\xi)}\epsilon^{\nu\alpha\beta}-\partial^\nu_{(\xi)}
\epsilon^{\mu\alpha\beta})\eqno(4.29)
$$
Inserting (4.29) in (4.28), we see that the second term vanishes because
$d^2\xi^\mu//d^2\eta^\alpha$.  The first term gives
$$
BPD=\lim_{\rho ,\delta ,m\rightarrow 0}-i\theta b^2 \soma \inte $$
 $$\times
\epsilon^{\nu\alpha\beta}\partial^\mu_{(\xi)}{\cal F}^{-1}[\frac{1}{
k^2+\theta^2}](\xi -\eta)\eqno(4.30)
$$
where we
used (4.16).

Writing
$$
\frac{1}{k^2+\theta^2}=\frac{1}{k^2}-\frac{\theta^2}{k^2(k^2+\theta^2)}
\eqno(4.31)
$$
and using the last equality in (2.32), we see that
the first term of (4.31) when inserted in (4.30) leads to an integral
identical to the one evaluated in Appendix D.  Hence, using (D.2) and
(4.31), we obtain
$$
BPD=-i\pi\theta b^2 \soma [arg(\vec x_i -\vec x _j)]
+K(T_x,T_y)\eqno(4.32a)
$$
or
$$
BPD=\lim_{\epsilon\rightarrow 0} i2\pi\theta b^2\ [arg(\vec x-\vec y)
+arg(\vec y-\vec x)-2 arg(\vec\epsilon)] +K(T_x,T_y)\eqno(4.32b)
$$
where
$$
K(T_x,T_y)=\limrd i\theta^3 b^2 \soma \inte
$$
$$\times
\partial^\mu_{(\xi)}
F(\xi -\eta)\eqno(4.33)
$$
where $F(\xi-\eta)={\cal F}^{-1}[{1\over k^2(k^2+\theta^2)}]$.

Observe that as the previous terms, also BPD contains a surface
dependent term, namely $K(T_x,T_y)$, given by (4.33).

Let us consider now the DCD term.  Using (4.21) and the $\theta$-dependent
part of (4.15), we find immediately
$$
DCD=\limrd \frac{-i\theta^3 b^2}{2}\soma \inte $$
 $$\times
\epsilon^{\mu\nu\sigma}\epsilon^{\alpha\beta\lambda}\epsilon
^{\sigma\rho\lambda}\partial_\rho^{(\xi)}F(\xi -\eta)\eqno(4.34)
$$
where
$F(\xi-\eta)={\cal F}^{-1}[{1\over k^2(k^2+\theta^2)}]$.  Observing that

$$
\epsilon^{\mu\nu\sigma}\epsilon^{\alpha\beta\lambda}\epsilon^{\sigma
\rho\lambda}\partial^{(\xi)}_\rho=\epsilon^{\alpha\beta\nu}\partial^\mu
_{(\xi)}-\epsilon^{\alpha\beta\mu}\partial^\nu_{(\xi)}\eqno(4.35)
$$
we see that the second term gives a vanishing contribution
because $d^2\xi^\mu//d^2\xi^\alpha$.  The first term is identical to
(4.33) and we get
$$
DCD=-\frac{1}{2}K(T_x,T_y)\eqno(4.36)
$$

Let us obtain, finally, the BCB term.  This is given by
$$
BCB=\limrd \frac{i\theta b^2}{2}\soma \inte $$
 $$\times
F^{\mu\nu}_{(\xi)}\ _\sigma F^{\alpha\beta}_{(\eta)} \ _\lambda
 \epsilon^{\sigma\rho\lambda}\partial^{(\xi)}_\rho F(\xi -\eta)
\eqno(4.37)
$$
where again $F(\xi-\eta)={\cal F}^{-1}[{1\over k^2(k^2+\theta^2)}]$.

We now have the identity
$$
F^{\mu\nu}_{(\xi)}\ _\sigma  F^{\alpha\beta}_{(\eta)}\ _\lambda
\epsilon^{\sigma\rho\lambda}\partial^{(\xi)}_\rho =\epsilon^{\nu\gamma\beta}
\partial^\mu_{(\xi)}\partial^\gamma_{(\xi)}\partial^\alpha_{(\eta)}+
\epsilon^{\mu\gamma\alpha}\partial^\nu_{(\xi)}\partial^\gamma_{(\xi)}
\partial^\beta_{(\eta)}- \epsilon^{\nu\gamma\alpha}\partial^\mu_{(\xi)}
\partial^\gamma_{(\xi)}\partial^\beta_{(\eta)}-$$
 $$-\epsilon^{\mu\gamma\beta}
\partial^\nu_{(\xi)}\partial^\gamma_{(\xi)}\partial^\alpha_{(\eta)}\eqno(4.38)
$$

Again, the contribution of the $2^{nd}$ term vanishes because $d^2\xi^\mu//d^2
\eta^\alpha$.  In Appendix E we show that the contribution of the $3^{rd}$
and $4^{th}$ terms also vanish.  We also show in Appendix E that the $1^{st}$
 term
gives
$$
BCB=-\frac{1}{2} BPD\eqno(4.39a)
$$
or
$$
BCB=\lim_{\epsilon\rightarrow 0}-i\pi\theta b^2[arg(\vec x-\vec y)+
arg(\vec y-\vec x)-2 arg(\vec
\epsilon)]-\frac{1}{2}K(T_x,T_y)\eqno(4.39b)
$$
Collecting the six terms given by (4.22), (4.26), (4.27), (4.32), (4.36)
and (4.39), we see that the surface dependent part of the last three
terms, namely $K(T_x,T_y)$ exactly cancel!  On the the other hand, we can
see, by using the fact that
$$
{\cal F}^{-1}[\frac{k^2}{k^2+\theta^2}]+\theta^2{\cal F}^{-1}[\frac{1}{
k^2+\theta^2}]={\cal F}^{-1}[1]=\delta^3(z-z')\eqno(4.40)
$$
that the sum of the surface
dependent pieces of the first three terms precisely add up to
$S_R[A_\mu]$ and therefore are exactly canceled by the renormalization
counterterm in (4.20)!  Just the surface independent part of the
above six terms contributes to the correlation function and we get:
 $$
<\mu_{\theta ,R}(x)\mu^*_{\theta ,R}(y)>=\exp\{\pi b^2[\frac{e^{-\theta
|x-y|}}{|x-y|}]+ 2\pi\theta ^2 b^2[\frac{1-e^{-\theta |x-y|}}
{|x-y|}]$$
$$
+i\pi\theta b^2[arg(\vec x-\vec y)+arg(\vec y-\vec x)]\}\eqno(4.41)
$$
where we defined the renormalized $\mu_\theta$ as
$$
\mu_{\theta ,R}=\lim_{\epsilon\rightarrow 0}\mu_\theta\exp [\frac{\pi
b^2}{2|\epsilon |}+\pi b^2\theta^2-i\pi\theta b^2arg(\vec \epsilon )]
\eqno(4.42)
$$
   Observe
that for $\theta\rightarrow 0$ (4.41) reduces to (2.37).  Eq. (4.41) is
multivalued up to a factor $\exp[i 2\pi(\pi\theta b^2)]$, indicating the
 $\mu_\theta$ has statistics $S_{\mu_\phi}=\pi\theta b^2$.  This is in
agreement with the previous result (4.9).

We see that $\mu_\theta$ is indeed a local field.  Notice that the
correlation function of $\mu$ would be given by $\exp[BPB+BCB]$,
indicating therefore that $\mu$ would be a nonlocal (surface dependent)
field in the MCS theory.  The same would be true for $\Sigma'$ (even in
Maxwell theory) whose correlation function would be given by
$\exp[DPD+DCD]$.  We see that the operators studied in [24] are nonlocal
in the MCS theory and therefore cannot be used as bona fide dual
operators.

The above study reveals how delicate and stringent are
the conditions necessary for the obtainment of local dual operators.

Observe that
$$
<\mu_{\theta R}(x)\mu_{\theta ,R}^*(y)> \tendei constant \neq 0
$$
 indicating that $\mu_\theta$ does not create genuine
vortex excitations in the MCS theory.  In section 5 we will see how this
result is changed when we add a Higgs potential to the MCS theory.

\bigskip
\leftline{\large\bf 4.5) Mixed and Composite Anyon Correlation Functions}
\bigskip
Let us study here the mixed $\sigma-\mu_\phi$ correlation function, from
which we will obtain the composite anyon $\varphi$ correlation function.

Combining (4.13) with (4.19), we can write .
$$
<\sigma(x_1)\mu_\theta(x_2)\sigma^*(y_1)\mu^*_\theta(y_2)>=\limrd
Z^{-1} \int DW_\mu \exp\{-\int d^3z[\frac{1}{2}W_\mu[P^{\mu\nu}+
\theta C^{\mu\nu}+G^{\mu\nu}]W_\nu
$$
$$+
C_\mu(z;x_1,y_1)W^\mu +W_\mu[P^{\mu\nu}+\theta C^{\mu\nu}]A_\nu(z;x_2,y_2)
+\frac{1}{4}A^2_{\mu\nu}(x_2,y_2)]\}\eqno(4.43)
$$

Integrating over $W_\mu$, we get
$$
<\sigma(x_1)\mu_\theta(x_2)\sigma^*(y_1)\mu^*_\theta(y_2)>-\limrd
\exp\{\frac{1}{2}\int d^3zd^3z'[C_\mu(z;x_1,y_1)+B_\mu(z;x_2,y_2)$$
$$
+D_\mu(z;x_2,y_2)]
[C_\nu(z';x_1,y_1)+B_\nu(z';x_2,y_2)+D_\nu(z';x_2,y_2)]$$
 $$ \times
[P^{\mu\nu}+\theta
C^{\mu\nu}+G^{\mu\nu}]^{-1}(z-z')-S_R[A_\mu(x_2,y_2)]\}\eqno(4.44)
$$
where $C_\mu$ was defined in
(2.39-40), $B_\mu$ in (2.28-29) and $D_\mu$ in (4.21).  Inserting
expression (4.15) for $[P^{\mu\nu}+\theta C^{\mu\nu}+G^{\mu\nu}]^{-1}$, we
will get, in addition to the six terms computed in (4.4) six more terms
involving $C_\mu$, namely CPC, CPB, and CPD, corresponding to the
$[P^{\mu\nu} +\partial^\mu\partial^\nu]$ part of (4.15) and CCC, CCB and
CCD, corresponding to the $\epsilon^{\mu\alpha\nu}$ part of (4.15).  As
before, we only consider the gauge invariant part of eventually gauge
dependent terms.

The CPC and CCC terms were computed in section 4.3.  Their sum is given
by the exponent in (4.17b).  The CPB term was computed in section 2.5.
It is given by (2.51).  The gauge invariant part of the CPD term is
given by
$$
CPD=\limrde ab\theta \soma \integ
$$
 $$\times
\partial^{(\eta)}_\beta arg(\vec\eta -\vec s_j)
\epsilon^{\alpha\beta\lambda}
[\delta^{i\lambda}\frac{e^{-\theta [|\xi -\eta |^2+|\epsilon |^2]
^{1/2}}}{4\pi[|\xi -\eta |^2+|\epsilon |^2]^{1/2}}]\eqno(4.45)
$$
 where we used (4.16), and introduced the
ultraviolet regulator $\epsilon$.
( We use the same convention for $r_i$ and $s_i$
as in section 2.5).  Using now (2.43a) and the results of Appendix C, we
find
$$
CPD=\lim_{\epsilon\rightarrow 0}-\frac{ab\theta}{4\pi}\soma (2\pi)^2
[-\frac{\exp [-\theta [|r_i -s_j|^2+|\epsilon |^2]^{1/2}]}{\theta}]
\eqno(4.46a)
$$
 or
$$
CPD=-\pi ab [e^{-\theta |x_1-y_2|}+e^{-\theta |x_2-y_1|}-e^{-\theta
|x_1-x_2|} -e^{-\theta |y_1-y_2|}]\eqno(4.46b)
$$

Let us consider now the CCD term.  This is given by
$$
CCD=\limrd iab\theta^2 \soma \integ
$$
 $$\times
\partial^{(\eta)}_\beta arg(\vec\eta -\vec s_j)
\epsilon^{\alpha\beta\lambda}
\epsilon^{i\rho\lambda}\partial^{(\xi)}_\rho F(\xi -\eta)\eqno(4.47)
$$
  where
$F(\xi-\eta) ={\cal F}^{-1}[{1\over k^2(k^2+\theta^2)}]$.
 In Appendix F, we show
that $CCD=0$.

Let us finally evaluate the CCB term, which is given by
$$
CCB=\limrd -ab\theta \soma \integ
$$
 $$\times
\partial^{(\eta)}_\beta arg(\vec\eta -\vec s_j)
 F^{\alpha\beta}_{(\eta)}\ _\lambda
\epsilon^{i\rho\lambda}\partial^{(\xi)}_\rho F(\xi -\eta) \eqno(4.48)
$$

where $F(\xi-\eta)$ is the same as in (4.47).  Observe now that
 $$
F^{\alpha\beta}_{(\eta)}\ _\lambda\epsilon^{i\rho\lambda}
\partial^{(\xi)}_\rho=\epsilon^{i\rho\beta}\partial^\alpha_{(\eta)}
\partial^\rho_{(\xi)}-\epsilon^{i\rho\alpha}\partial^\beta_{(\eta)}
\partial^\rho_{(\xi)}\eqno(4.49)
$$
As is shown in Appendix C, the only possibility for the
$\xi$-integral in (4.48) to be nonzero is the expression between
brackets (in (4.48)) being contracted with a derivative $\partial^i$.
We could try to force the appearance of such a derivative by using an
identity such as (2.43).  Such identity, however, would also be
contracted with (4.49) and we immediately see that the derivative part
of the identity would vanish, because of the contraction with the
$\epsilon$'s and $\partial_\rho$.  We conclude, therefore, that
$CCB=0$.

Collecting all terms contributing to the mixed correlation function, we
get
$$
<\sigma(x_1)\mu_{\theta ,R}(x_2)\sigma^*(y_1)\mu_{\theta ,R}(y_2)>=
\exp\{ \frac{\pi a^2}{\theta}[e^{-\theta |x_1-y_1|}-1]+$$
$$
\pi b^2[\frac{e^{-\theta |x_2-y_2|}}{|x_2-y_2|}]+2\pi \theta^2 b^2
[\frac{1-e^{-\theta |x_2-y_2|}}{|x_2-y_2|}]-\pi b[e^{-\theta |x_1-y_2|}
+e^{-\theta |x_2-y_1|} -e^{-\theta |x_1-x_2|}-e^{-\theta |y_1-y_2|}]$$
$$+ i\pi ab[arg(\vec x_1-\vec x_2)+ arg(\vec y_1-\vec y_2)- arg(\vec
x_1-\vec y_2)-arg (\vec y_1 -\vec x_2)]$$
$$
+i\pi\theta b^2[arg(\vec x_2-\vec y_2)+arg(\vec y_2 -\vec x_2)]\}
\eqno(4.50)
$$

Observe that this expression reduces to (2.52) when $\theta\rightarrow 0$.
The correlation function for the composite anyon field (2.53) can now be
easily obtained by taking the limit $x_1\rightarrow x_2$ and
$y_1\rightarrow y_2$ in (4.50).  Introducing the same renormalization
factor as in  (2.53), we get
$$
<\varphi(x)\varphi^*(y)>=\exp\{[\frac{\pi a^2}{\theta}-2\pi ab]
[e^{-\theta |x-y|}-1]+\pi b^2[\frac{e^{-\theta |x-y|}}{|x-y|}]
$$
$$
+2\pi\theta ^2b^2[\frac{1-e^{-\theta |x-y|}}{|x-y|}]-i\pi(ab-\theta
b^2)[arg(\vec x-\vec y)+arg(\vec y-\vec x)]\}\eqno(4.51)
$$
   Notice that the multivaluedness of
$<\varphi\varphi ^*>$ corresponds to a phase $\exp[i 2\pi(\pi
ab-\pi\theta b^2)]$.  This is in agreement with (4.11) and indicates that
$\varphi$ has statistics $S_\varphi =\pi b|a-\theta b|$.  Observe that for
$\theta\rightarrow 0$, (4.51) reduces to (2.54), which we obtained in
Maxwell theory.

\bigskip
\leftline{\Large\bf 5) The Maxwell-Chern-Simons-Higgs Theory}
\bigskip
\leftline{\large\bf 5.1) The Unbroken Phase}

Let us study now the Maxwell-Chern-Simons-Higgs (MCSH) theory which is
essentially the Abelian Higgs Model plus a Chern-Simons term:
$$
{\cal L}=-\frac{1}{4}W^2_{\mu\nu}+\frac{\theta}{2}\epsilon
^{\mu\alpha\beta}W_\mu \partial_\alpha W_\beta +|D_\mu \phi |^2
-g^2 \phi^*\phi -\frac{\lambda}{4}(\phi^*\phi)^2 \eqno(5.1)
$$
As in the AHM, we have two phases, according to whether: i)
$g^2>0$, $<\phi>=0$ or ii) $g^2<0$, $<\phi>=\kappa \neq 0$.
Shifting the Higgs field around $\kappa$ (for $g^2<0$), as before, will
generate a mass term $(M=e\kappa)$ for $W_\mu$.

Let us introduce here the operators $\sigma$ and $\mu_\theta$ considered
before.  The equal time commutation rules of the MCSH theory are the
same as in (4.5) and therefore all the commutators evaluated in section
4.2 remain valid here.

The same remarks concerning the surface independence of $\mu$ in the AHM
may be applied to $\mu_\theta$ here and therefore we conclude that (2.7)
is the appropriate surface renormalization counterterm for $\mu_\theta$ in
the MCSH theory too.

Let us consider firstly the symmetric phase ($g^2 > 0$,
$<\phi>=0$).  The $\mu_\theta$-correlation function is given by
$$
<\mu_\theta(x)\mu^*_\theta(y)>\limrd Z^{-1}\int DW_\mu \exp\{-\int
d^3z[\frac{1}{2}W_\mu[P^{\mu\nu}+\theta C^{\mu\nu}+G^{\mu\nu}]W_\nu+
|D_\mu \phi |^2+
$$
$$ V(\phi)+W_\mu[P^{\mu\nu}+ \theta C^{\mu\nu}]A_\nu(z;x,y)+\frac{1}{4}
A^2_{\mu\nu}(z;,x,y)]\}\eqno(5.2)
$$

As in the AHM, $<\mu_\theta\mu^*_\theta>$ will be given by (3.3).  Again,
only two-leg graphs will contribute to the large distance limit in (3.3)
[17].  Also here, we are going to make a loop expansion for the
computation of $\Lambda(x,y)$ in (3.3).  As in the AHM, the leading
contribution will be given by the graph of Fig. 4b.  The only difference
now is that the $W_\mu$-propagator is given by (4.15) and the
external field is $B_\mu(z;x,y)+D_\mu (z;x,y)$, where $B_\mu$ is given by
(2.28) and $D_\mu$ by (4.21).  We immediately see that the graph of Fig.
4b, for these external fields and propagator is identical to the first
term in the exponent in (4.20).  It follows that
$$
\lim_{|x-y|\rightarrow\infty} <\mu_{\theta ,R}(x)\mu^*_{\theta ,R}(y)>
_{MCSH}=<\mu_{\theta ,R}(x)\mu^*_{\theta ,R}(y)>_{MCS} \tendei
constant \neq 0 \eqno(5.3)
$$
 where
 $<\mu_{\theta ,R}\mu^*_{\theta ,R}>_{MCS}$ is given by (4.41) and we
renormalized $\mu_\theta$ as in the MCS theory.

We conclude that also in the ordered phase of the MCSH theory the
$\mu_\theta$ operator does not create genuine vortex excitations.  We will
see that this is no longer true in the broken phase of the theory.

Let us evaluate now the order correlation function $<\sigma\sigma^*>$.
This is given now by
$$
\cors \limrd Z^{-1}\int DW_\mu \exp\{-\int d^3z[\frac{1}{2} W_\mu[P^{\mu
\nu}+\theta C^{\mu\nu}+G^{\mu\nu}]W_\nu+ |D_\mu \phi |^2 $$
$$+
V(\phi)+ C_\mu(z;x,y)W^\mu ]\}\eqno(5.4)
$$
 where $C_\mu(z;x,y)$ was defined in eqs.
(2.39) and (2.40).  As in the AHM, $<\sigma\sigma^*>$ is given by (3.6).
Making exactly the same approximation we did in section 3.2 we see that
$\Gamma(x,y)$ will be given by the graph in Fig. 4b but with $C_\mu$ as
the external field and (4.15) as the $W_\mu$-propagator.  It is easy to
see that this is given by the exponent in (4.14).  We readily conclude
that in lowest order, the long distance behavior of the gauge invariant
part of $<\sigma\sigma^*>$ in the MCSH theory is given by (4.17),
namely,
$$
\lim_{|x-y|\rightarrow\infty}<\sigma_R(x)\sigma^*_R(y)>_{MCSH}=
<\sigma_R(x)\sigma_R^*(y)>_{MCS} \tendei 1
\eqno(5.5)
$$
where $<\sigma_R\sigma^*_R>_{MCS}$ is given by (4.17c) and
we renormalized $\sigma$ as in the MCS theory.

We now see that the behavior of $<\sigma_R\sigma^*_R>$ indicates charge
screening in spite of the fact that we are in the symmetric phase.  This
naturally happens, because as in the MCS theory, the $W_\mu$- field
acquired a mass through the Chern-Simons mechanism.

As before, the long distance behavior of the composite anyon field could
be obtained by just exchanging $B_\mu(z;x,y)+D_\mu(z;x,y)$ by
$B_\mu(z;x,y)+C_\mu(z;x,y)+D_\mu(z;x,y)$ in (5.2).  We would obtain (4.51) for
the long distance behavior of $<\varphi\varphi^*>$.

\bigskip
\leftline{\large\bf 5.2) The Broken Phase}

Let us investigate now the broken phase, where $<\phi>\not= 0$.
Shifting the Higgs field around the vacuum value, as before, we can see
that a mass term will be generated for $W_\mu$.  The quadractic
lagrangian, however, contains a Chern-Simons term in addition to (3.8).
The euclidean $W_\mu$- propagator will be now given by
$$
D_{\mu\nu}(k)=\frac{P^{\mu\nu}(1+\frac{M^2}{k^2})-\theta\epsilon^{
\mu\alpha\nu}k_\alpha}{(k^2+M^2)^2+k^2\theta^2} +\frac{k^\mu k^\nu}
{k^2(\xi k^2+M^2)} \eqno(5.6)
$$
Observe that this reduces to (3.9a) for $\theta\rightarrow 0$ and to
(4.15a) for $M\rightarrow 0$.

Let us evaluate the order correlation function $<\sigma\sigma^*>$.  This
is given by (5.4) and (3.6).  Making the same approximation as before,
the leading contribution to the long distance behavior of
$<\sigma\sigma^*>$ will be given by the graph of Fig. 4b with $C_\mu$
as the external field which, as we pointed out before, leads to an
expression similar to (4.14).  The only difference is that now we must
use the $M\not= 0$ and $\theta\not= 0$ propagator (5.6) instead of
(4.15).  As in the previous cases we are going to compute the gauge
independent part of $<\sigma\sigma^*>$, namely the $\delta^{\mu\nu}$
and $\epsilon^{\mu\alpha\nu}$ proportional terms.  As in the MCS theory,
the $\epsilon^{\mu\alpha\nu}$-proportional term vanishes (see Appendix
C).  The $\delta^{\mu\nu}$-proportional contribution, namely
$$
\delta^{\mu\nu}{\cal F}^{-1}[\frac{k^2+M^2}{(k^2+M^2)^2+k^2\theta^2}]=
\frac{\delta^{\mu\nu}}{2\pi ^2}\int^\infty_0 dk\,k \frac{\sin k|x|}{|x|}\,
[\frac{k^2+M^2}{(k^2+M^2)^2+k^2\theta^2}] \eqno(5.7)
$$
can be calculated with the help of the identity
$$
\delta^{\mu\nu}\ \frac{\sin k|x|}{|x|}=\partial^\mu\partial^\nu
[-\frac{\cos k|x|}{k}]+\frac{x^\mu x^\nu}{|x|^2}\ \  [\, \frac{\sin k|x|}
{|x|} - k \cos k|x|\, ]\eqno(5.8)
$$
which generalizes (2.43) when inserted in (5.7).  Proceeding exactly as
we did to go from (2.42) to (2.44), we see that the contribution from
the last two terms in (5.8) vanishes.  The contribution from the first
term gives
$$
\cors \tendei \exp \{4\pi^2 a^2[I(x-y)-I(0)]\}\eqno(5.9a)
$$
 or
$$
<\sigma_R(x)\sigma^*_R(y)> \tendei \exp\{4\pi^2a^2 I(x-y)\}\eqno(5.9b)
$$
  where
$$
I(x-y)=\frac{1}{2\pi^2} \int^\infty_0 dk \cos k|x-y|\, [\frac{k^2+M^2}{
(k^2+M^2)^2+k^2\theta^2}]\eqno(5.10)
$$
and
$\sigma_R\equiv \sigma \exp[2\pi^2 a^2 I(0)]$. Observe that
$$
\lim_{\theta\rightarrow 0}I(x-y)=\frac{e^{-M|x-y|}}{4\pi M}\eqno(5.11)
$$
 and (5.9) reduces to (3.11) in the AHM.

It follows from the Riemann-Lebesgue lemma that $\lim_{|x-y|\rightarrow
\infty} I(x-y)= 0$.  Hence,
$<\sigma_R(x)\sigma^*_R(y)> \tendei 1$
.  This result indicates that also in
the broken phase of the MCSH theory the charge is screened.  Actually it
is doubly screened, both by M and by $\theta$, namely by the Higgs
mechanism and by the Chern-Simons term.  Notice that for $M\rightarrow
0$ and $\theta\rightarrow 0$, we have
$$
\lim_{M,\theta\rightarrow
0}I(x-y)=\frac{1}{4\pi}[\frac{1}{M}-|x-y|]\eqno(5.12)
$$
and (5.9) reduces to the
expression (3.7) found in the symmetric phase of the AHM, where charge
screening no longer occurs.

Let us consider now the disorder correlation function
$<\mu_\theta\mu^*_\theta>$.  According to (5.2), we can write
$<\mu_\theta\mu^*_\theta>$ exactly as in (3.12), with the only
difference that now $S[W_{\mu\nu},\phi,D_\mu\phi]$ is the action
corresponding to (5.1).  Making, as before, the change of variable
$W_\mu\rightarrow W_\mu+A_\mu(z;x;y)$, we get an expression identical to
(3.13), where again, S is associated with (5.1).  The correlation
function $<\mu_\theta\mu^*_\theta>$ therefore, will still be given by
(3.14), the only difference being that the $W_\mu$- propagator used in
the computation of Feynman graphs must be (5.6) instead of (3.9a).  As
before, making an expansion in loops and in powers of $M^2$, it follows
that the leading contribution to $\tilde \Lambda(x,y)$ is given by the
graphs of Fig. 6.  Using (5.6), it is easy to see that the sum of the
gauge dependent graphs of Fig. 6a identically vanishes for all values of
the gauge parameter $\xi$.  The first contribution (of order $0({M^2\over
\hbar}))$ is given, as in the AHM, by the graphs of Fig. 6b.  These are
independent of $\theta$ and we see that in lowest order the
$\mu_{\theta,R}$-correlation function behaves asymptotically at large
distances exactly as (3.18).  The $\theta$-dependence will only be
introduced by 1-loop corrections, through the $W_\mu$-propagator.

The result $<\mu_{\theta ,R}(x)\mu^*_{\theta ,R}(y)>\tendei 0$
 in the broken phase of the MCSH theory shows that
$\mu_{\theta,R}$ creates true topological charge bearing excitations in
this case.  The mass of these quantum solitons is $M_v=\pi M^2b^2$,
within our approximation.

Let us remark finally that the (charge and magnetic flux bearing)
composite anyon operator $\varphi$ would be the creation operator of the
quantum excitations corresponding to the classical electrically
charged vortices of ref. [25].

\bigskip
\leftline{\Large\bf 6) Conclusions and Remarks}

The order-disorder duality structure was exploited in order to introduce
dual operators corresponding to a local U(1) symmetry and carrying
respectively, charge and magnetic flux, in some $2+1$ dimensional field
theories involving a vector field.  The conditions for the obtainment
of local order($\sigma$), disorder (vortex) ($\mu$) or anyon ($\varphi$)
operators must be carefully analyzed in each case.  Local, surface
independent correlation functions of these operators can be obtained
only after some very stringent requirements are met.

In the absence of a Chern-Simons term, namely, in the Maxwell and in the
Abelian-Higgs theories, the behavior of $<\sigma>$ and $<\mu>$ which can
be infered from the long distance behavior of the $\sigma$ and $\mu$
correlation functions is analogous to the behavior of the corresponding
operators in the Ising model.  The Maxwell and symmetric Higgs phases
correspond to the disordered phase of the latter and the broken Higgs
phase corresponds to the ordered phase of it.  The ordered phase
contains vortex excitations whose mass is explicitly evaluated.  The
disordered phase may be viewed a vortex condensate.  Massive anyon
excitations occur in all cases.  The screening of charge in the broken
phase is clearly exposed by the behavior of the $\sigma$-correlation
function.

In the presence of a Chern-Simons term, namely, in the
Maxwell-Chern-Simons and Maxwell-Chern-Simons-Higgs theories, we have
two concurrent mechanisms of charge screening (mass generation for the
vector field).  The behavior of $\sigma$-correlation functions expresses
the Chern-Simons induced charge screening in the MCS and symmetric MCSH
phases.  In the broken MCSH phase we can infer from the long distance
behavior of $<\sigma\sigma^*>$ the action of both mechanisms of charge
screening, namely, the Chern-Simons and Higgs mechanisms

Massive vortex excitations are still only seen to occur in the broken
phase of the MCSH theory.  We therefore can clearly distinguish the
process which generates a mass for the vector field from that which
generates a mass for the vortices.  The first one can occur either via
the Chern-Simons or Higgs mechanisms, while the second one is only
induced by the latter.

In the presence of a Chern-Simons term we show that the statistics is no
longer determined by the product $Q\Phi$ (charge $\times$  flux) but rather, by
$\tilde Q\Phi$, where
$\tilde Q=Q-\theta\Phi$.  As a consequence, the exclusively flux bearing
disorder (vortex) operator $\mu_\theta$ is seen to be anyonic.

The charge and magnetic flux carrying quantum vortex excitations
corresponding to the classic solutions of the MCSH theory found in [25]
will be described by the composite anyon operator
$\varphi=\sigma\mu_\phi$.

Let us remark that even in the presence of a Chern-Simons term, an
Ising-like behavior for the order and disorder variables could be
obtained, by introducing a new order operator $\sigma '=\sigma\phi$,
where $\phi$ is the Higgs field.  This could be done naturally in the
Abelian Higgs and MCSH theories but could also be easily achieved in the
Maxwell and MCS theories by considering the weak coupling limit of them
with a complex scalar field (without self-interaction, for instance).
The behavior of $<\sigma '>$ in the MCS and MCSH symmetric and broken
phases would be the same as that of $<\sigma>$ in the Maxwell and
Abelian Higgs (symmetric and broken) phases.  Furthermore, the short
distance behavior of $<\sigma '\sigma '^*>$ in the Maxwell and MCS
theories would have the singularity one would expect usually. The
main advantage of the operator $\sigma$ is that it provides
a reliable order parameter which works well
 whether or not a Higgs field is present
in the theory.

\vfill
\newpage
\leftline{ \Large\bf Acknowledgements}

I would like thank K.D. Rothe for his constructive criticism in the
early stages of this work.  I am very grateful to the Physics Department
of Princeton University and especially to C. Callan and D.Gross  for the the
kind hospitality.

This work is supported in part by CNPq- Brazilian National Research
Council .

\bigskip
\bigskip
\bigskip
\leftline{\Large\bf Appendix A}

Let us show here the Cauchy-Riemann equations involving the real and
imaginary parts of the analytic function $\ln (z-x)$ that we use in this
work.

The first one is (we will be in euclidean space throughout this
Appendix)
$$
\partial_iarg(\vec z-\vec x)=-\epsilon^{ij}\partial_j\ln |\vec z-\vec
x| + 2\pi \int ^\infty_{x,L} d\xi_j\epsilon^{ij} \delta^2(z-\xi)
\,\,\,,\,\,\, i=1,2 \eqno(A.1)
$$

The line integral corresponds to the singularity of the derivative of
$arg(\vec z-\vec x)$ along its cut $L$, chosen to be the straight line
going from $\vec x$ to $\infty$ along the $z^1$ axis.

{}From (A.1) we get a second equation by contracting with $\epsilon^{ij}$:
$$
\epsilon^{ij}\partial_i arg(\vec z-\vec x)=-\partial^j\ln |\vec z-\vec
x| +2\pi \int^\infty_{x,L} d\xi^j \delta^2(z-\xi)\,\,\,,\,\,\,i=1,2
\eqno(A.2)
$$

{}From (A.2) we obtain

$$
\epsilon^{ij}\partial_i\partial_jarg(\vec z -\vec x)=2\pi \delta^2(z-x)
+ 2\pi \int^\infty_{x,L} d\xi^i \partial^{(\xi)}_i \delta^2(z-\xi)
\eqno(A.3)
$$
where we used the fact that $\partial_i\partial^i\ln |\vec z-\vec x|=
2\pi \delta^2(z-x)$.

\bigskip
\leftline{\Large\bf Appendix B}

Let us derive here eqs. (2.13) and (4.6).  We start with (2.6) and observe
that
$$
A_{\mu\nu}=\partial_\mu A_\nu -\partial_\nu A_\mu =\int_{T_x(C)}
d^2\xi_\nu arg(\vec \xi-\vec x) \partial_\mu \delta^3(z- \xi)-
(\mu\leftrightarrow\nu)$$
$$
+\oint_{C(x)} d\xi^\alpha \epsilon_{\mu\nu\alpha} arg(\vec \xi -\vec x)
\delta^3(z-\xi)\eqno(B.1)
$$
In this equation, the second term comes from the discontinuity of $A_\mu$
at $C(x)$, the boundary of $T_x(C)$ (see Fig. 1).  Inserting (B.1) in
(2.6), we get
$$
\mu(x;C)=\exp\{ib \int_{T_x(C)} d^2\xi_\nu \partial_\mu W^{\mu\nu}
arg(\vec\xi -\vec x) -\frac{ib}{2}\oint _{C(x)} d\xi^\alpha\epsilon
_{\mu\nu\alpha}W^{\mu\nu}arg(\vec \xi-\vec x)\}\eqno(B.2)
$$
Observing that $\partial_\mu W^{\mu\nu}=\epsilon_{\mu\sigma\rho}
\epsilon^{\nu\lambda\rho} \partial^\sigma \partial_\lambda W^\mu$ and
integrating by parts the first term in (B.2), with the help of Stokes
theorem we see that the boundary term exactly cancels the last term in
(B.2).  The remaining term gives
$$
\mu(x;C)=\exp\{-ib \int_{T_x(C)} d^2\xi_\nu\, W^{\mu\nu}\partial_\mu
arg(\vec \xi- \vec x)\}\eqno(B.3)
$$
Noting that $d^2\xi_\nu$ only has the zeroth component and remembering
that $E^i=W^{io}$, eq. (2.13) immediately follows.

In order to obtain eq. (4.6) for $\mu_\theta$, let us consider $\Sigma'$,
eq. (2.10b).  Inserting (B.1) in (2.10b) (we can write this equation in
terms of $A_{\mu\nu}$ divided by two) we get
$$
\Sigma '(x;C)=\exp\{ia\int_{T_x(C)} d^2\xi_\mu \epsilon^{\mu\alpha\beta}
\partial_\alpha W_\beta arg(\vec \xi-\vec x)+ia \oint_{C(x)} d\xi_\mu
W^\mu arg(\vec \xi -\vec x)\}\eqno(B.4)
$$
Integrating by parts the first term with the help of Stokes theorem we
again find that the boundary term exactly cancels the last term. The
remaining term yields the following expression:
$$
\Sigma '(x;C)=\exp\{-ia\int_{T_x(C)} d^2\xi_\mu\epsilon^{\mu\alpha\beta}
W_\beta \partial_\alpha arg(\vec \xi-\vec x)\}\eqno(B.5)
$$
 This
equation, for $a=\theta b$ together with (B.3) leads to expression (4.6)
for $\mu_\theta$.

\bigskip
\leftline{\Large\bf Appendix C}

Let us demonstrate here the following two useful results:
$$
R_1=\limrd \int_{R_x(C)} d^2\xi[\epsilon^{ij}\partial_i arg(\vec \xi-
\vec x)+\partial_j \ln |\vec \xi-\vec x|]\partial_j\Lambda(x^0,\vec \xi)
=2\pi \Lambda(x^0,\vec x)\eqno(C.1)
$$
 and
$$
R_2=\limrd \int_{T_x(C)} d^2\xi \epsilon^{ij}\partial_i arg(\vec\xi-
\vec x)\partial_j \Lambda(x^0,\vec x)=2\pi\Lambda(x^0,\vec x)\eqno(C.2)
$$
for an arbitrary function $\Lambda(x)$.

  Let us start with (C.1).  We may write $R_1$ as (we ommited the
argument $x^0$)
$$
R_1=\limrd \int_{R_x(C)} d^2\xi [\epsilon^{ij}\partial_j[\partial_i
arg(\vec\xi-\vec x)\Lambda(\vec\xi)]-\epsilon^{ij}\partial_j
\partial_i arg(\vec\xi -\vec x)\Lambda(\vec\xi)$$
$$
+\partial_i[\partial_i\ln |\vec\xi-\vec x|\Lambda(\vec\xi)]
-\partial_i\partial_i\ln |\vec\xi -\vec x|\Lambda(\vec\xi)]\eqno(C.3)
$$
Using (A.2) or (A.3), we see that the sum of the $2^{nd}$ and $4^{th}$ terms
is:
 $$
2^{nd} +4^{th}=\limrd 2\pi \int_{R_x(C)} d^2\xi \int^\infty_{x,L}
d\eta^j \partial^{(\eta)}_j \delta^2(\xi -\eta) \Lambda(\vec\xi)
\eqno(C.4)
$$
 $$
\,\,\,\,\,\,\,\,\,=\limrd 2\pi \int_{x,L\cap R_x(C)} d\eta_j
\partial^{(\eta)}_j\Lambda(\vec\eta)=\lim_{\rho\rightarrow 0}
2\pi[ \Lambda(\vec x +\rho \hat{\xi}_1) -\Lambda(\vec x)]=0
\eqno(C.5)
$$
where the last integral is performed over the section of $L$ contained
in $R_x(C)$.  Observe that (A.3) is not zero in a cut plane like
$R_x(C)$.

It would also be different from zero in a punctured plane, for instance
(this is the case considered in [11]).

The $1^{st}$ and $3^{rd}$ terms in (C.3) may be evaluated with the help of the
Stokes and (two dimensional) Gauss theorems, respectively.  The result
is
$$
R_1=\limrd \{-\oint_{C_\infty +C_\delta +C_\rho} d\xi^i [\partial_i
arg(\vec\xi -\vec x) +\epsilon^{ij}\partial_j\ln |\vec\xi -\vec x|]
\Lambda(\vec\xi)\}\eqno(C.6)
$$
 In this expression, $C_\rho$ is the arc of circumference
of radius $\rho$ and $C_\delta$, the two straight lines, in Fig. 3.
$C_\infty$ is the arc of circumference with infinite radius closing $C$
at infinity.  We immediately see that the contribution from $C_\delta$
is zero because the integrand is orthogonal to the integration element
along $C_\delta$.  The contribution of $C_\infty$ also vanishes because
of (A.1).  The only contribution to (C.6) comes from $C_\rho$.
Inserting (A.1) in (C.6) we get ,
$$
R_1=\lim_{\rho\rightarrow 0}2\pi \Lambda(x^0,\vec x+\rho \hat{\xi}_1)
=2\pi\Lambda(x^0,\vec x)\eqno(C.7)
$$
thus establishing (C.1).

Let us observe that replacing $\partial_j\Lambda$ by a vector $\Lambda_j$ (not
a derivative!) in (C.1) would yield a zero result.  Indeed, inserting
(A.2) in (C.1) we would obtain an integral identical to (C.5), with
$\partial_j\Lambda$ replaced by $\Lambda_j$.  This would vanish at
$\rho\rightarrow 0$ for any regular $\Lambda_j$.  The boundary terms, eq.
(C.6) which are only present when $\Lambda_j=\partial_j \Lambda$ are
responsible for making $R_1\not= 0$.  This is the reason why expressions
of the type we found in the last terms of (2.43) and (5.8) do not
contribute to correlation functions.

Let us turn now to (C.2).  Observe that $\epsilon^{ij}\partial_j
\partial_i arg(\vec\xi-\vec x)=0$ over the surface $T_x(C))$, Fig. 1.
Using this fact, we can make the $\partial_j$ derivative total in (C.2)
and then use Stokes theorem to get
$$
R_2=\limrd \oint_{C_\infty +C_\delta +C_\rho} d\xi^i \partial_i
arg(\vec\xi -\vec x) \eqno(C.8)
$$
Here $C_\rho$ is the arc of circumference of radius $\rho$ and
$C_\delta$, the straight lines in Fig. 1.  $C_\infty$ is the infinite
radius arc of circumference closing $C$ at infinity.  The contribution
from $C_\delta$ again vanishes because the integrand is orthogonal to
$d\xi^i$ along it.  The contribution from $C_\infty$ vanishes for
$\Lambda(\vec\xi)$-functions going to zero at infinity.  This is true for
all cases considered in this work.  Even for the function between square
brackets in (2.43b) this is seen to be valid by adopting the following
procedure before taking the limit $x\rightarrow\infty$: we put the
system inside a circular box of radius $R\equiv{1\over m}$.  We
immediately see that the above mentioned function will vanish on the
boundary of this box.  Then we make $R\rightarrow\infty$ as
$x\rightarrow\infty$.

The only nonvanishing contribution to (C.8) therefore comes from
$C_\rho$.  Making a Taylor expansion in $\Lambda(\vec\xi)$ around
$\vec\xi =\vec x$, we see that in the limit $\rho\rightarrow 0$ only the
zeroth order term is nonvanishing and we obtain:
$$
R_2=\limrd \Lambda(x^0,\vec x) \int_{C_\rho} d\xi^i\partial_i
arg(\vec \xi-\vec x)= 2\pi \Lambda(x^0,\vec x)\eqno(C.9)
$$
 This
establishes (C.2).

In eq. (4.7b), we have an expression similar to (C.2), the difference
being that the boundary of the integration region now contains two
curves: $C_x$ and $C_y$.  Using the Stokes theorem with $\partial_j$ as
the total derivative in $C_x$ and with $\partial_i$ in $C_y$, eq. (4.8)
immediately follows from (C.2).

\bigskip
\leftline{\Large\bf Appendix D}

Let us demonstrate here the following two useful results:
$$
R_3\equiv\limrdxy \int_{R_y(C)} d^2\eta [\epsilon^{ki}
\partial^{(\eta)}_karg(\vec\eta -\vec y)+\partial^{(\eta)}_i
\ln |\vec\eta -\vec y|]
$$
 $$ \times
\int_{T_x(C)}d^2\xi_\mu \partial^{(\xi)}_i
arg(\vec\xi -\vec x)\partial^\mu_{(\xi)}[\frac{1}{4\pi |\xi -\eta|}]
=\pi arg(\vec y-\vec x)\eqno(D.1)
$$
and
$$
R_4\equiv\limrdxy \int_{T_y(C)} d^2\eta_\alpha \epsilon^{\alpha\beta\nu}
\partial^{(\eta)}_\beta arg(\vec \eta -\vec y) \int_{T_x(C)} d^2\xi_\mu
\partial^{(\xi)}_\nu arg(\vec\xi -\vec x)\partial^\mu_{(\xi)}
[\frac{1}{4\pi|\xi -\eta |}]$$
$$
\,\,\,\,\,\,\,\,=\pi [arg(\vec y -\vec x)+ arg(\vec x-\vec y)]\eqno(D.2)
$$

Let us first demonstrate the following result that we will need later
$$
I_1\equiv \limrdx \int_{T_x(C)} d^2\xi_\mu arg(\vec\xi -\vec x)
\partial^\mu_{(\xi)} [\frac{1}{4\pi |\xi -\eta |}]=\frac{1}{2}
[arg(\vec\eta -\vec x) +\pi]\eqno(D.3)
$$
It may be written as
$$
I_1=\limrdx \int^{2\pi -\delta_x}_{\delta_x}d\varphi\,\varphi
\int^\infty_{\rho_x}dr\,r H[L^2+r^2-2Dr\cos(\varphi-\varphi_0)]
^{-3/2}\eqno(D.4)
$$
where $L=|\eta -x|$,
$\varphi_0=arg(\vec\eta-\vec x)$, $H= L \ \cos\theta$ and $D=L \ \sin\theta$,
where $\theta$ is the angle the 3-vector $\eta-x$ makes with the
$\xi_3$-axis in 3D-euclidean space.

Evaluating the r-integral [26], we obtain
$$
I_1=\lim_{\delta_x\rightarrow 0}\int^{2\pi -\delta_x}_{\delta_x}
d\varphi\,\varphi \frac{H}{L-D\cos(\varphi -\varphi_0)}\eqno(D.5)
$$

Performing now the $\varphi$-integral [27], we get the last equality in
(D.3).

Let us consider now the last integral in (D.1) and (D.2) (observe that
$\nu\equiv i=1,2$ in (D.2), because $d^2\eta^\alpha$ has only the
3-component different from zero.  Also $\mu\equiv 3$ in (D.1) and (D.2)
for similar reason):
$$
I_2\equiv\limrdx \int_{T_x(C)} d^2\xi \partial^{(\xi)}_i arg(\vec\xi -
\vec x)\partial^3_{(\xi)}[\frac{1}{4\pi |\xi -\eta |}]\eqno(D.6)
$$

Integrating by parts and using the $2D$-Gauss theorem, we may write
(D.6) as
$$
I_2=\limrdx\{\partial^{(\eta)}_i \int_{T_x(C)} d^2\xi_\mu arg(\vec\xi
-\vec x) \partial^\mu_{(\xi)}[\frac{1}{4\pi |\xi -\eta |}]+
$$
$$
\oint_{C(x)}d\xi^j \epsilon^{ji} arg(\vec\xi -\vec x)
\partial^3_{(\xi)}[\frac{1}{4\pi |\xi -\eta |}]\}\eqno(D.7)
$$
Using (D.3) in the first term of (D.7), inserting the result in (D.1)
and (D.2) and using (C.1) and (C.2), respectively, we see that the
first term of (D.7) contributes a term $\pi arg(\vec y-\vec x)$ to
(D.1)and (D.2).

We will now show that the second term (boundary term) in (D.7) only
contributes to (D.2).  Indeed, using (A.2) we see that in the limit
$\delta_x\rightarrow 0$, the first expression between brackets in (D.1) is
orthogonal to $\epsilon^{ji} d\xi^j$ along the two straight lines in
$C(x)$.  The contribution of the arc of circumference with radius $\rho$
vanishes for $\rho\rightarrow 0$.  Hence, only the first term of (D.7)
contributes to (D.1) and the result of the previous paragraph
establishes (D.1).

Let us compute now the contribution from the second term in (D.7) to
(D.2). Inserting this term in (D.2), we obtain the boundary contribution
to $R_4$:
$$
R^{(b)}_4=\limrdxy \oint d\xi^j arg(\vec\xi -\vec x) \int_{T_y(C)}
d^2\eta \partial^{(\eta)}_j arg(\vec\eta -\vec x)(-)\partial^3_{(\eta)}
[\frac{1}{4\pi |\xi -\eta |}]\eqno(D.8)
$$
Observe that the second integral again has an expression like (D.6).
Inserting  (D.7) in (D.8) we see that the second term vanishes because
$d\xi^j$ is orthogonal to $\epsilon^{kj}d\eta^k$ along the straight
lines of $C_x$ and $C_y$. As before, the contribution from the arcs of
circumference vanish for $\rho_x,\rho_y\rightarrow 0$.  Using (D.3) we
see that the first term of (D.7) gives the following contribution to
(D.8):
$$
R_4^{(b)}=-\frac{1}{2}\limrdx \oint_{C_x} d\xi^j arg(\vec\xi -\vec x)
\partial^{(\xi)}_j arg(\vec\xi -\vec y)$$
$$
\,\,=\frac{1}{2}\limrdx \oint_{C_x} d\xi^j[\partial^{(\xi)}_j
arg(\vec\xi -\vec x)] arg(\vec\xi -\vec y)=\pi arg(\vec x-\vec y)
\eqno(D.9)
$$
 where we used (C.9).  Combining this term with the
contribution from the first term of (D.7) to $R_4$, we immediately
establish the last equality in (D.2).

Let us finally mention that actually a more general result could be
established for (D.1) and (D.2).  Before taking the limit
$\rho_x,\delta_x\rightarrow 0$ in (D.1) one would obtain $R_3= {1\over
2} arg(\vec y-\vec x)\Omega (\vec y,T_x(C))$, where $\Omega(\vec
y,T_x())$ is the solid angle determined by the point $\vec y$ and the
surface $T_x(C)$.  For $\rho_x,\delta_x\rightarrow 0\ \ , \Omega (\vec
y,T_x(C))\rightarrow 2\pi$ and we recover (D.1).  Also, without taking
the limits in (D.2) one would obtain $R_4={1\over 2}[arg(\vec y -\vec
x)\Omega(\vec y,T_x(C)) + arg(\vec x-\vec y)\Omega (\vec x,T_y(C))]$,
which reduces to (D.2) for $\rho_x,\delta _x\rightarrow 0$ and
$\rho_y,\delta_y\rightarrow 0$.  These results would be relevant for
the computation of correlation functions of the loop dependent operators
$\mu(x;C)$ (finite $\rho,\delta$).
\vfill
\newpage

\leftline{\Large\bf Appendix E}

Let us demonstrate here the result (4.39) for the BCB term  which
is given by eq. (4.37).

We start showing that the contribution from the last two terms in (4.38)
vanishes.  Insertion of the $3^{rd}$ term of (4.38) in (4.37) yields an
integral
$$
H(x_i,x_j)=\limrd \inte
$$
 $$\times
 \epsilon^{\nu\gamma\alpha}\partial^\mu_{(\xi)}
\partial^\gamma_{(\xi)}\partial^\beta_{(\eta)} F(\xi -\eta)\eqno(E.1)
$$
Making the change of variable $\xi\leftrightarrow \eta$ and exchanging
$\alpha\leftrightarrow\mu$, $\beta\leftrightarrow\nu$, and using the
fact that $\partial_{(\xi)} F(\xi -\eta) = -\partial_{(\eta)}
F(\xi-\eta)$, we immediately conclude that the $4^{th}$ term in (4.38) would
yield an integral $H(x_j,x_i)$.  Hence, summing over $i$ and $j$ in (4.37)
we see that the contributions from the last two terms in  (4.38) are
identical.  Let us now show that these contributions actually vanish.
Using the fact that $d^2\xi^\mu$ and $d^2\eta^\alpha$ only have the
3-component different from zero and noting that $arg(\vec \eta -\vec
x_j)=0$ along the surface $T_{x_j}(C)$, we may write
$$
H(x_i,x_j)=\limrd \int_{T_{x_i}(C)} d^2\xi \epsilon^{kl} \partial^{(\xi)}_k
arg(\vec\xi -\vec x_i)
$$
$$\times
 \partial^l_{(\xi)}\partial^3_{(\xi)}
\int _{T_{x_j}(C)} d^2\eta \partial^{(\eta)}_i [\partial^{(\eta)}_i
arg(\vec\eta -\vec x_j) F(\xi -\eta)]\eqno(E.2)
$$
Using now the Stokes and 2D-Gauss theorems, we get
$$
H(x_i,x_j)=\limrd \oint _{C(x_i)} d\xi^k \partial^{(\xi)}_k
arg(\vec\xi -\vec x_i) \oint _{C(x_j)} d\eta^j \epsilon^{ji}
\partial^{(\eta)}_i arg(\vec\eta -\vec x_j)[\partial^3_{(\xi)}
F(\xi -\eta)]\eqno(E.3a)$$
$$
\,\,\,\,\,\,\,\,=\limrd 2\pi \oint _{C(x_j)} d\eta^j \partial^{(\eta)}_j
\ln |\vec\eta -\vec x_j| [\partial^3_{(x_i)}F(x_i -\eta)]\eqno(E.3b)
$$
 where we used (C.9) and (A.2).

The contribution from the arc of circumference of radius $\rho$ (Fig. 1)
to (E.3b) vanishes because $d\eta^j$ is orthogonal to the integrand
along this curve.  The contributions from each one of the straight lines
in $C(x_j)$ (Fig.1) cancel each other in the limit
$\delta_{x_j}\rightarrow 0$ because the integrand is regular along the
cut.  Hence we conclude that $H(x_i,x_j)=0$.

The only nonvanishing contribution to (4.37) comes from the first term
in (4.38).  Insertion of this term in (4.37) yields the following
integral
$$
J(x_i,x_j)=\limrd \inte
$$
$$\times
 \epsilon^{\nu\gamma\beta}\partial^\mu_{(\xi)}
\partial^\gamma_{(\xi)}\partial^\alpha_{(\eta)} F(\xi -\eta)\eqno(E.4)
$$
where $F(\xi-\eta)={\cal F}^{-1}[{1\over
k^2(k^2+\theta^2)}]$.

Before evaluating (E.4), let us demonstrate the following results that
we will need later:
$$
\int _{T_x(C)} d^2\xi\  \partial_\mu arg(\vec\xi -\vec x)\frac{(\xi -
y)^\mu}{|\xi -y|} f(|\xi -y|)=0\eqno(E.5a)
$$
and
$$
\int _{T_x(C)} d^2\xi\  \partial_\nu \partial_\mu arg(\vec\xi -\vec
x)\frac{(\xi -
y)^\mu}{|\xi -y|} f(|\xi -y|)=0\eqno(E.5b)
$$
where $f$ is an arbitrary function.

We may write (E.5a) as
$$
\int^\infty_\rho dr\, r\int^{2\pi-\delta}_\delta d\varphi \frac{1}{r}
[\frac{D\sin(\varphi -\varphi_0)}{[D^2+r^2-2Dr\cos(\varphi -\varphi_0)]
^{1/2}}] f([L^2+r^2-2Dr\cos(\varphi -\varphi_0)]^{1/2})\eqno(E.6)
$$
where $L,D$ and $\varphi_0$ were
defined in Appendix D.  The angular integration in (E.6) is easily seen
to vanish, because the contribution from the first two quadrants exactly
cancels the one coming from the last two.  This establishes (E.5a). Eq.
(E.5b) immediately follows if we note that it will be given by an
expression like (E.6) but with ${1\over 2}$ replaced by
$-{\delta^{\nu r}\over r^2}$.

Let us now return to (E.4).  Observe that $\partial^\alpha_{(\eta)}$
only has the 3-component different from zero (because it is contracted
with $d^2\eta^\alpha$).  Since $arg(\vec \eta-\vec x_j)$ does not depend
on $\eta^3$, we may make $\partial^\alpha_{(\eta)}$ a total derivative
and use Gauss theorem on the closed surface $S$, consisting of
$T_{x_j}(C) \cup T_\infty(C)\cup \widetilde T_{x_j}(C)$. Here $T_\infty(C)$ is
a copy of $T_{x_j}(C)$ but at $\eta^3=\infty$ and $\widetilde T_{x_j}(C)$ is
the surface paralell to the $\eta^3$-axis, connecting $T_{x_j}(C)$ and
$T_\infty(C)$, which is obtained by translating the curve $C(x_j)$ from
$\eta^3=x^3_j$ to $\eta^3=\infty$ along the $\eta^3$-axis.  For
$\rho,\delta\rightarrow 0$ of course, $\widetilde T_{x_j}(C)$ collapses to
zero.  After using the Gauss theorem we may write (E.4) as,
$$
J(x_i,x_j)=\limrd \int_{T_{x_i}(C)} d^2\xi_\mu \int_{V(S)} d^3\eta\,
\epsilon^{\nu\gamma\beta} \partial^{(\xi)}_\nu arg(\vec\xi -\vec x_i)
\partial^{(\eta)}_\beta arg(\vec\eta -\vec x_j)$$
 $$
\,\,\,\,\, \times\partial^\mu_{(\xi)}\partial^\gamma_{(\xi)}[-\Box_\eta
F(\xi -\eta)]\eqno(E.7)
$$
where $V(S)$ is the volume bounded by $S$ (note that $d^2\eta^\alpha$
points inwards $V(S)$). To arrive at (E.7), we used (E.5b) and the fact
that $\Box arg(\vec\eta-\vec x_j) =0$ on the volume $V(S)$.  Observe that the
contribution from $T_\infty(C)$ vanishes because of the boundary
conditions at infinity we impose on the function $F$ (see Appendix C).
One may also show that the contribution from $\widetilde T_{x_j}(C)$ vanishes
for $\rho,\delta\rightarrow 0$ as should be expected.

Observe now that $\partial^\gamma_{(\xi)}$ also only has the 3-component
different from zero in (E.7).  This is so because it is contracted with
$\epsilon^{\nu\gamma\beta}$ and $\nu ,\beta=1,2$ since these indexes are
contracted with derivatives of arg functions.  We therefore can pull
$\partial^\gamma$ as a total derivative in (E.7) (after using the fact
that $\partial^\gamma_{(\xi)} =-\partial^\gamma_{(\eta)}$) in the same
way we did with $\partial^\alpha_{(\eta)}$ in (E.4) and use Gauss theorem
backwards to obtain
$$
J(x_i,x_j)=\limrd \inte
$$
$$\times
 \epsilon^{\nu\alpha\beta}\partial^\mu_{(\xi)}
[-\Box F(\xi -\eta)]\eqno(E.8)
$$
Again we used the fact that the contribution from $T_\infty(C)$ is zero
for the same reason above.  The contribution from $\widetilde T_{x_j}(C)$
also vanishes because the integrand in (E.8) is orthogonal to
$d^2\eta^\alpha$ along $\widetilde T_{x_j}(C)$.

The integral $J(x_i,x_j)$ in (E.8) is identical to the one in (4.30)
($-\Box F={\cal F}^{-1}[(k^2+\theta^2)^{-1}]$), therefore, insertion of (E.8)
in
(4.37) immediately demonstrates (4.39).

\bigskip
\leftline{\Large\bf Appendix F}

Let us demonstrate here that (4.47) is equal to zero.  Using the
identity
$\epsilon^{\alpha\beta\lambda}\epsilon^{i\rho\lambda}=\delta^{\alpha
i}\delta^{\beta\rho}-\delta^{\alpha\rho}\delta^{\beta i}$, we see
that the first term vanishes because $d^2\eta^\alpha$ is orthogonal to
the expression between square brackets in (4.47).  The second term, when
inserted in (4.47) yields the integral
$$
L(x_i,x_j)=\limrd (-) \integ
$$
$$\times
\partial^{(\eta)}_i arg(\vec\eta -\vec s_j)
\partial^\alpha_{(\xi)}F(\xi
-\eta)\eqno(F.1)
$$
Observing that $\partial^\alpha_{(\xi)}=-\partial^\alpha_{(\eta)}$ only
has the 3-component different from zero and using the Gauss theorem
exactly in the same way we did in Appendix E, we may write (F.1) as
 $$
L(x_i,x_j)=\limrd \integra
$$
$$\times
\partial^{(\eta)}_i arg(\vec\eta -\vec s_j)
 [-\Box_\eta F(\xi -\eta)] \eqno(F.2)
$$
where $V(S)$ is the volume introduced in Appendix E.

As we have shown in Appendix C, the only possibility for the
$\xi$-integral to give a nonzero result is the $\eta$-integral in (F.2)
producing a total derivative.  Using the identity given by (2.43a), for
$[-\Box F]$ in (F.2), we see that the $\eta$-integral of the derivative term
vanishes according to (E.5a).  Hence we conclude that
$\lim_{\rho_{r_i},\delta_{r_i}\rightarrow 0}L(x_i,x_j)= 0 $ and
therefore $CCD=0$, in eq. (A.47).

It is worth mentioning that for the special case $F={\cal F}^{-1}[{1\over
k^2}]$, $-\Box F=\delta^3(\xi-\eta)$ in (F.2) and then, we would obtain
$L(x_i,x_j)=2\pi arg(\vec r_i-\vec s_j)$.  This, however is not the case
of the theories studied here.

\bigskip
\bigskip
\bigskip
\leftline{\Large\bf References}
\medskip

\begin{description}
\item[1-]
S. Deser, R. Jackiw and S. Templeton, {\it Ann. of Phys.} {\bf 140} (1982) 372;
{\it Phys. Rev. Lett.} {\bf 48} (1982) 975; W. Siegel, {\it Nucl. Phys.}
{\bf
B156} (1979) 135; J. Schonfeld, {\it Nucl. Phys.} {\bf B185} (1981) 157; R.
Jackiw and S. Templeton, {\it Phys. Rev.} {\bf D23} (1981) 2291;

\item[2-]
F. Wilczek and A. Zee, {\it Phys. Rev. Lett.} {\bf 51} (1983) 2250; A.M.
Polyakov, {\it Mod. Phys. Lett.} {\bf A3} (1988) 325

\item[3-]
F. Wilczek, {\it Phys. Rev. Lett.}
{\bf 48} (1982) 1144; {\it Phys. Rev. Lett.} {\bf
49} (1982) 957

\item[4-]
E.C. Marino and J.A. Swieca, {\it Nucl. Phys.} {\bf B170} [FS1] (1980) 175;

\item[5-]
R. Prange and S. Girvin, {\it ``The Quantum Hall Effect''}, Springer-Verlag,
New York, (1990).

\item[6-]
V. Kalmeyer and R.B. Laughlin, {\it Phys. Rev. Lett.} {\bf 59} (1987) 2095;
R.B. Laughlin, {\it Phys. Rev. Lett.} {\bf 60} (1988) 2677; A. Fetter, C. Hanna
and R.B. Laughlin, {\it Phys. Rev.} {\bf B39} (1989) 9679; Y.H. Chen, F.
Wilczek, E. Witten and B.I. Halperin, {\it Int. J. of Mod. Phys.} {\bf B3}
(1989) 1001;  T. Banks and J.D. Lykken, {\it Nucl. Phys.} {\bf B336} (1990) 500

\item[7-]
F. Wilczek, {\it ``Fractional Statistics and Anyons''}, World Scientific,
Singapore, (1990).

\item[8-]
V. Kurak and J.A. Swieca, {\it Phys. Lett.} {\bf B82} (1979) 289;  R. K\"oberle
and J.A. Swieca, {\it Phys. Lett.} {\bf B86} (1979) 209;  R. K\"oberle, V.
Kurak and J.A. Swieca, {\it Phys. Rev.} {\bf D20} (1979) 897

\item[9-]
E. Fradkin and L.P. Kadanoff, {\it Nucl. Phys.} {\bf B170} [FS1] (1980) 1

\item[10-]
L.P. Kadanoff and H. Ceva, {\it Phys. Rev.} {\bf B3} (1971) 3918

\item[11-]
E.C. Marino, B. Schroer, J.A. Swieca, {\it Nucl. Phys.} {\bf B200} [FS4]
(1982) 473
\item[12-]
E.C. Marino, {\it ``Dual Quantization of Solitons''}, in Proceedings of the
NATO
Advanced Study Institute {\it ``Applications of Statistical and Field Theory
Methods in Condensed Matter''}, D. Baeriswyl, A. Bishop and J. Carmelo,
eds., Plenum, New York, (1992)

\item[13-]
J. Fr\"ohlich, {\it Commun. Math. Phys.} {\bf 47} (1976) 269; J. Fr\"ohlich,
in {\it ``Recent Developments in Gauge Theories''} (Carg\`{e}se, 79), G.'t
Hooft et
al, eds. Plenum, New York, (1980); J. Fr\"ohlich and P.A. Marchetti,
{\it Commun. Math. Phys.} {\bf 116} (1988) 127

\item[14-]
E.C. Marino, {\it Phys. Rev.} {\bf D38} (1988) 3194

\item[15-]
J. Fr\"ohlich and P.A. Marchetti, {\it Commun. Math. Phys.} {\bf 112} (1987)
343

\item[16-]
J. Fr\"ohlich and P.A. Marchetti, {\it Lett. Math. Phys.} {\bf 16} (1988) 347;
{\it Commun. Math. Phys.} {\bf 121} (1989) 177; J. Fr\"ohlich, F. Gabbiani and
P.A. Marchetti, in Proceedings of the Banff Summer School in Theoretical
Physics, H.C. Lee, ed., to appear.

\item[17-]
G.W. Semenoff and P. Sodano,{\it Nucl. Phys.} {\bf B328} (989) 753; M.
L\"uscher, {\it Nucl. Phys.} {\bf B326} (1989) 557; R. Jackiw and S.Y. Pi,
{\it Phys. Rev.} {\bf D42} (1990) 3500;
A.Kovner, B. Rosenstein and D. Eliezer, {\it Nucl. Phys.} {\bf B350} (1991)
325; {\it Mod. Phys.Lett.} {\bf A5} (1990) 2733; V.F. M\"uller, {\it Z. Phys.}
 {\bf
C51} (1991) 665

\item[18-]
E.C. Marino, {\it Phys. Lett.} {\bf B263} (1991) 63; E.C. Marino, {\it
``Bosonic
Representations of Fermions and Anyons in $2+1D$''}, to appear.

\item[19-]
S.C. Zhang, T.H. Hansson and S. Kivelson, {\it Phys. Rev. Lett.} {\bf 62}
(1989) 82; N. Read, {\it Phys. Rev. Lett.} {\bf 62} (1989) 86

\item[20-]
E. Fradkin, {\it ``Field Theories of Condensed Matter Systems''},
Addison-Wesley, Redwood City (1991);

\item[21-]
E.C. Marino, G.C. Marques, R.O. Ramos and J. Stephany-Ruiz, {\it Phys. Rev.}
{\bf D45} (1992) 3690

\item[22-]
H.B. Nielsen and P. Olesen, {\it Nucl. Phys.} {\bf B61} (1973) 45

\item[23-]
H. de Vega and F. Schaposhik, {\it Phys. Rev.} {\bf D14} (1976) 1100

\item[24-]
A. Leclair, {\it Phys.Lett.} {\bf B264} (1991) 355

\item[25-]
S. Paul and A. Khare, {\it Phys. Lett.} {\bf B174} (1986) 420

\item[26-]
I.S. Gradshteyn and I.M. Ryzhik, {\it ``Table of Integrals, Series and
Products''}, Academic Press, New York (1980)

\item[27-]
A.P. Prudnikov, Yu.A. Brichkov and O.I. Marichev,{\it ``Integrals and
Series''}, Gordon and Breach,  New York (1988).

\end{description}

\vfill
\newpage
\leftline{\Large\bf Figure Captions}
\bigskip
\noindent Fig. 1: Surface used in the definition of $\mu(x;C)$.

\noindent Fig. 2:
Surface used in the definition of $\sigma(x)$.

\noindent Fig. 3:
Surfaces used in the definition of the external field
$A_\mu(x;C)$.

\noindent Fig. 4a: Vertex involving the external fields $B_\mu, C_\mu$
or $D_\mu$.

\noindent Fig. 4b: Leading graph contributing to the long distance
behavior of correlation functions in the AHM and MCSH theory.

\noindent Fig. 5: Vertices relevant for the evaluation of $<\mu\mu^*>$
in the broken phase of the AHM and MCSH theory: a) gauge independent; b)
gauge dependent.

\noindent Fig. 6: Leading graphs contributing to the long distance
behavior of $<\mu\mu^*>$ in the broken phase of the AHM and MCSH theory:
a) gauge dependent; b) gauge independent.

\end{document}